\pdfoutput=1
 \documentclass[twocolumn]{autartar} 
\usepackage{graphicx} % for pdf, bitmapped graphics files
\usepackage{epsfig} % for postscript graphics files
\usepackage{amsmath} % assumes amsmath package installed
\usepackage{amssymb}  % assumes amsmath package installed
\usepackage{float}
\usepackage{color}
\usepackage{bm}
\usepackage{enumerate}
 \usepackage{algorithm}
 \usepackage{algorithmicx}
 \usepackage{algpseudocode}
 \usepackage{booktabs}
\usepackage{mathtools}
\usepackage{harvard}
\newtheorem{defi}{Definition}

\newtheorem{lemm}{Lemma}
\newtheorem{rema}{Remark}
\newtheorem{ass}{Assumption}
\newcommand{\asys}{f}

\newcommand{\flow}{\psi} %this is used as the flow of a dynamical system

\newcommand{\xdim}{n}

\renewcommand{\P}{P}
\newcommand{\varvar}{\sigma^2}
\newcommand{\mom}{M}

\newcommand{\momp}{\mom_{1..\P}}
\newcommand{\cmomp}{\bar{\mom}_{1..\P}}

\newcommand{\xini}{x_{\text{ini}}}

\newcommand{\invset}{\mathcal{I}}
\newcommand{\probdist}{\lambda}

\newcommand{\roaact}{\mathcal{R}}
\newcommand{\roa}{\bar{\mathcal{R}}}
\newcommand{\roaacttrue}{\mathcal{R}^{\ast}}
\newcommand{\roatrue}{\bar{\mathcal{R}}^{\ast}}
\newcommand{\roazero}{\roaact_0}

\newcommand{\R}{\mathbb{R}}
\newcommand{\Rn}{\mathbb{R}^n}

\newcommand{\pdeg}{\partial} %polynomial degree

\newcommand{\trans}{^T}
\newcommand{\innz}{\in\mathbb{N}_0}
\newcommand{\rr}{r} %polynomial degree
\newcommand{\sls}{\Omega}

\newcommand{\vsls}{\rho}

\newcommand{\slsvr}{\sls_{\lyp_{\vsls}}}
\newcommand{\slsvriso}{\sls_{\lyp_{\vsls=1}}}
\newcommand{\slsvro}{\sls_{\lyp_{1}}}
\newcommand{\pcesys}{\bar{f}}
\newcommand{\p}{p} %order of expansion
\newcommand{\basisg}{\Phi} %general orthogonal polynomial
\newcommand{\normfac}{\gamma}

\newcommand{\meas}{\mu} %measure (probability)
\newcommand{\sspace}{\Theta} %sample space

\newcommand{\spar}{a}
\newcommand{\cspar}{\bar{\spar}}
\newcommand{\spart}{c}
\newcommand{\cspart}{\bar{\spart}}
\newcommand{\spartt}{c}

\newcommand{\spartwo}{y}
\newcommand{\cspartwo}{\bar{\spartwo}}

\newcommand{\cx}{\bar{x}}
\newcommand{\cxdot}{\dot{\bar{x}}}
\newcommand{\cz}{\bar{z}}

\newcommand{\pring}{\mathcal{P}}
\newcommand{\pvariate}{n}
\newcommand{\ranvar}{\xi}
\newcommand{\ltwo}{\mathcal{L}_2}

\newcommand{\gt}{T} %Galerkin tensor

\newcommand{\cxini}{\bar{x}_{\text{ini}}}
\newcommand{\cxep}{\cx_{\text{EP}}}
\newcommand{\xep}{x_{\text{EP}}}
\newcommand{\lyp}{V}
\newcommand{\dlyp}{\dot{V}}

\newcommand{\Qlyp}{Q_{\lyp}}

\newcommand{\mb}{s_1}
\newcommand{\mc}{s_2}

\newcommand{\la}{l}
\newcommand{\lb}{l}

\newcommand{\sos}{\Sigma}
\newcommand{\sosc}{\Sigma[\cx]}

\newcommand{\ballset}{\mathcal{B}} %sublevel set of inside ball
\newcommand{\ballmat}{B} %gram matrix of ball
\newcommand{\insideball}{b} %inside ball polynomial
\newcommand{\mean}{m}

\newcommand{\roamat}{Q_0}
\newcommand{\varfix}{\hat{\sigma}^2}

\def\bary{
	\begin{bmatrix}
	\bar{y}_0\\
		\bar{y}_J
\end{bmatrix}}

\begin{document}

\begin{frontmatter}



\title{Region of attraction analysis of nonlinear stochastic systems using Polynomial Chaos Expansion}


\author{Eva Ahbe, Andrea Iannelli, Roy S. Smith}

\address{Automatic Control Laboratory, Swiss Federal Institute of Technology (ETH Zurich), Physikstrasse 3, 8092 Zurich, Switzerland, {\tt\small $\{$ahbee$,$iannelli$,$rsmith$\}$@control.ee.ethz.ch}. }

\begin{abstract}
		A method is presented to estimate the region of attraction (ROA) of stochastic systems with finite second moment and uncertainty-dependent equilibria. The approach employs Polynomial Chaos (PC) expansions to represent the stochastic system by a higher-dimensional set of deterministic equations. We first show how the equilibrium point of the deterministic formulation provides the stochastic moments of an uncertainty-dependent equilibrium point of the stochastic system. 
	A connection between the boundedness of the moments of the stochastic system and the Lyapunov stability of its PC expansion is then derived. Defining corresponding notions of a ROA for both system representations, we show how this connection can be leveraged to recover an estimate of the ROA of the stochastic system from the ROA of the PC expanded system.	
	Two optimization programs, obtained from sum-of-squares programming techniques, are provided to compute inner estimates of the ROA. The first optimization program uses the Lyapunov stability arguments to return an estimate of the ROA of the PC expansion. Based on this result and user specifications on the moments for the initial conditions, the second one employs the shown connection to provide the corresponding ROA of the stochastic system. The method is demonstrated by two examples.
\end{abstract}




\end{frontmatter}


\section{Introduction}
The analysis of the region of attraction (ROA) of an uncertain nonlinear system is an active field of research \cite{Chesi2004,Valmorbida2017,Iannelli2019}. 
The type of uncertainty and its appearance in the dynamical equations is often pivotal for the choice of the analytical approach. A class of uncertain systems commonly considered has two characteristic properties: firstly, the equilibrium point of the system is independent of the uncertainty, and secondly, the uncertainty comes from a uniform distribution over a finite range of values. The stability of this class of systems can be analysed using Lyapunov methods where an estimate of the ROA is obtained in the form of the sublevel set of a Lyapunov function. The aim then lies in finding a Lyapunov function verifying a largest possible estimate of the ROA. For systems where the uncertainty itself is parametric and polytope-bounded, parameter-dependent as well as common and composite Lyapunov functions have been investigated in, e$.$g$.$, \citeasnoun{Topcu2010}, \citeasnoun{Chesi2004}, \citeasnoun{Iannelli2019}. While estimates for these cases can be efficiently obtained, the assumption of uncertainty-independent equilibria and uniformly distributed uncertainty excludes most systems from the analysis as equilibria are in general uncertainty-dependent  and the stochasticity affecting the system can come from a wide range of distributions.

The ROA analysis in the case of uncertainty-dependent equilibria is not directly amenable to the use of Lyapunov functions, as this method requires knowledge of the equilibrium's location in the standard case. To tackle this problem, an equilibrium-independent version of the ROA was proposed in \citeasnoun{Iannelli2018} where the idea is to formulate the ROA as a function of a new coordinate representing the deviation of the state relative to the equilibrium point. This approach, however, is still limited to uncertainties from uniform distributions. A more general approach for stability analysis is provided by contraction methods which inherently do not require knowledge on the equilibrium state. Contraction of uncertain systems was studied, e$.$g$.$ in \citeasnoun{Ahbe2018b} for polytope-bounded parametric uncertainty and in \citeasnoun{Bouvrie2019}, \citeasnoun{Pham2009} for It$\hat{\text{o}}$ stochastic differential equations. Contraction methods often pose, however, numerically more complex problems compared to Lyapunov analysis as they consider the differential system. Furthermore, while contraction analysis gives conclusions about the contractive behaviour of a system it in general does not provide information on the state of the (stochastic) equilibrium.

In this work we present an efficient method to analyse the ROA of stochastic nonlinear systems with uncertainty-dependent equilibrium points where the uncertainty can be in form of any square-integrable random variable or process. The stochastic system is thereby represented by a higher-dimensional set of deterministic equations obtained from a Polynomial Chaos (PC) expansion of the stochastic dynamics. PC expansions are a polynomial approximation method which allow the representation of a second order process, i$.$e$.$ stochastic systems with finite second moment, by a higher-dimensional deterministic expression. An overview of PC expansions can be found, e$.$g$.$, in \citeasnoun{Sullivan2015} and \citeasnoun{Lemaitre2010}. While PC expansion techniques have become established tools in uncertainty quantification, their use in stability and control is still sparse \cite{Kim2013a} and mostly focused on linear systems. Stability analysis of linear stochastic systems via PC expansions using Lyapunov inequalities was previously performed in \citeasnoun{Fisher2009} and \citeasnoun{Lucia2017}. In \citeasnoun{Hover2006}, the evolution of the stochastic modes resulting from the PC expansion was used to obtain information on the stability of a nonlinear system. A more generalized approach for polynomial systems using Lyapunov arguments is briefly presented in \citeasnoun{Fisher2008}, however the method proposed therein can only be used to certify global stability properties. 

This paper proposes a novel method to analyse the ROA of stochastic nonlinear systems with uncertainty dependent equilibria by leveraging the PC expansion framework.
We first show how an equilibrium point of the deterministic expression given by the PC expansion corresponds to an uncertainty-dependent equilibrium point of the stochastic system. The latter can be represented as a set, which we refer to as the \textit{equilibrium set}, for which statistical information is directly obtained from the expansion coefficients. 
For both the stochastic system and its PC expansion notions of local stability are provided, consisting in boundedness of moments for the first and asymptotic stability in the sense of Lyapunov for the second. It is then demonstrated how Lyapunov stability of the PC expansion equilibrium point implies moment boundedness of trajectories in the neighborhood of the equilibrium set of the stochastic system. From the stability notions and their shown connection, corresponding notions of the ROA are defined for both system representation. To obtain an inner estimate of the ROA of the PC expanded system, Lyapunov arguments stating sufficient conditions are formulated and converted into an algorithm. The algorithm employs well-established sum-of-squares verification techniques to test polynomial positivity \cite{Parrilo2000} which were previously used for analysing the ROA of polynomial systems in, e$.$g$.$ \citeasnoun{JarvisWloszek2005}, \citeasnoun{Topcu2008} and others.
We then proceed by providing a notion of the ROA of the stochastic system which is formulated on the basis of the ROA of its deterministic PC expansion. While the ROA of a deterministic system is clearly defined, the definition of an attractive region of uncertain system can be of various types. For stochastic systems a definition of the ROA can be derived from the type of stochastic stability under consideration. For an overview of the different definitions of stochastic stability see, e$.$g$.$, \citeasnoun{Khasminskii2012}. A widely used notion for the ROA of uncertain systems is that of a `robust' ROA, which is the intersection of the ROA's obtained for each realization of the uncertainty. As it thus relates to the worst case, this notion is suitable for uncertainties with uniform distributions but less so for other distributions where the worst case is not of practical interest or exploiting the statistical information available gives less conservative results.
A probabilistic ROA of an uncertainty-independent equilibrium point was investigated for Ito-stochastic system via Lyapunov functions in \citeasnoun{Gudmundsson2018}. In \citeasnoun{Steinhardt2012} `safe sets' of a controlled system with quantified failure probabilities were considered and computed with a supermartingale approach. We here provide an approach in which the ROA is obtained in terms of the region of initial conditions with specified moment properties for which trajectories almost surely converge to the equilibrium set of the stochastic system. The moment properties of the initial condition consist of, for example, a fixed variance in the initial state and can be specified by the user. The proposed method is demonstrated by two examples from the literature.

The paper is structured as follows. In Section 2 the problem statement and the method of PC expansion is introduced. The notions of stability of stochastic systems under consideration are presented in Section 3 and the connection of the stability concepts between the stochastic system and its PC expansion are shown. Lyapunov conditions for the local stability and the analysis methods to obtain an estimate of the ROA of the PC expanded system including the formulation of the corresponding optimization problem are given in Section 4. Further, the connection of the ROA of the PC expanded system to the ROA of the stochastic system and the computation of the latter under user-defined moment specifications are presented. The examples are shown in Section 5 and a conclusion is presented in Section 6.

\subsection{Notation}

Let $(\sspace,\mathcal{F},\meas)$ be a probability space, where $\sspace$ is a sample space, $\mathcal{F}$ is a $\sigma$-algebra of the subsets in $\sspace$ and $\meas$ is a probability measure on $(\sspace,\mathcal{F})$. 
The Lebesgue space is denoted by $\mathcal{L}_l$, where $0\leq l\leq\infty$.
% (Definition in Sullivan p.17). 
The inner product in the $\ltwo$ space is denoted by $\langle \cdot, \cdot \rangle_{\ltwo(\meas)}$ which represents integration (i.e. expectation) with respect to $\mu$. Expectation is further indicated by $\mathbb{E}$.
A random variable $\ranvar: \sspace \rightarrow \R$ with finite second moment, $\ranvar \in \mathcal{L}_2(\sspace,\meas)$, is referred to as the stochastic germ. For clarity of presentation we take the stochastic germ $\ranvar$ to be one-dimensional in this work. The extension to vector valued $\ranvar$ with independent components is straightforward, see e$.$g$.$ \citeasnoun{Sullivan2015}.
Let the $\P$-th moment of a random variable $\ranvar$ be given by $\mom_{\P}(\ranvar)=\mathbb{E}[|\ranvar|^\P]$.
A probability distribution $\probdist$ with $\P$ given moments, where $1\leq\P<\infty$, is denoted by $\probdist(\momp)$. The symbol $\sim$ denotes an element with distribution $\probdist$.

Let $\pring^\pvariate$ denote the ring of all $\pvariate$-variate polynomials with real coefficients and let $\pring^\pvariate_{\leq \rr}$ denote those polynomials of total degree at most $\rr \in \mathbb{N}_0$.
A polynomial $g(x):\Rn \rightarrow \R, g(x) \in \pring^\pvariate_{\leq \rr}$ is called a sum-of-squares (SOS) if it can be written as $g(x) = \sum_{i} q_i(x)^2, \, q_i(x) \in \pring^n_{\leq \rr/2} $. Moreover, $g$ is SOS if and only if there is a matrix $Q \succeq 0$ such that  $g(x) = v(x)\trans Qv(x)$, where $v(x)$ is a vector of monomials. The set of all SOS polynomials in the indeterminant $x$ is indicated by $\sos[x]$.  
The degree of a polynomial $g$ in $x$ is indicated by $\pdeg(g)$.

\section{Problem Statement and Background}
\label{sec:system}
In this work we are interested in estimating the region of attraction of the equilibrium state of a stochastic nonlinear system. 
%For deterministic systems, the ROA is defined as the set of initial conditions from which trajectories converge to the equilibrium of a system. 

The systems we consider are continuous time second order processes of the form
\begin{equation}
\label{eq:sys}
\dot{x}(t,\ranvar) = \asys(x(t, \ranvar),\spar(\ranvar)),
\end{equation}
where $x\in \Rn$ is the random state variable, $\spar: \R \rightarrow \R^m$ is an independent random variable and $\asys: \Rn \times \R^m \rightarrow \Rn$ is assumed to be polynomial in $x$ and $\spar$.  
%For the higher-dimensional case as well as details on probability spaces, see, e$.$g$.$ \cite{Sullivan2015}.
%which is possible for the class of systems we consider here through the Weierstrass Theorem \cite{Weierstrass1885}.

We consider systems with an uncertainty-dependent attractive equilibrium point $\xep(\ranvar)$. Let the set, given by the evaluation of $\xep(\ranvar)$ for each realization of the uncertainty, be denoted by
\begin{equation}
\label{eq:invset}
\invset = \{ x \in \Rn \, | \, \asys(x(\ranvar),\spar(\ranvar)\!)\! =\!0, \,\,\ranvar\! \in\! \mathcal{L}_2(\sspace,\meas) \}.
\end{equation}
In the following, the set $\invset$ of a system is referred to as the \textit{equilibrium set}.
%The set consisting of the uncertainty-dependent equilibrium points $x_{\text{EP}}(\spar)$  of a stochastic system \eqref{eq:sys} is denoted by $\invset = \{x_{\text{EP}}(\spar(\ranvar))\}_{\ranvar \in \ltwo}$. Due to the continuity of \eqref{eq:sys} in $\spar$, the set $\invset$ is connected.
 %Note that for notational simplicity we here consider the same probability distribution for the initial state as for the independent random variable. 

Let $\psi(t, \xini(\ranvar),\ranvar)$ denote the uncertainty-dependent solution of \eqref{eq:sys} at time t with initial condition $\xini(\ranvar)$, where the initial state is also allowed to be random, i$.$e$.$ $x(t=0) = \xini(\ranvar)$.
The ROA of the equilibrium set $\invset$ is then defined as 
%\begin{equation}
%\label{eq:stochroaset}
%%\roa = \{x_\text{ini} \in \Rn | \,\,\mathbb{P}[\lim\limits_{\,\,t \rightarrow \infty} \text{dist}(\psi(t, \xini),\invset) \rightarrow 0] = 1	\,\,\, \text{a.s.}\}
%%&\!\!\!\!\roaacttrue = \{\xini \in \Rn | \\[2pt]\label{eq:stochroaset}
%%& ||\mathbb{E}[|\psi(t, \spar(\ranvar), \xini (\ranvar))|^{\P}] - \mathbb{E}[|\invset|^{\P}]||_2 \xrightarrow[\,\,t \rightarrow \infty] {} 0\}
%%&\mathbb{P}[\lim\limits_{\,\,t \rightarrow \infty} d(\psi(t, \xini (\ranvar)),\invset) \rightarrow 0] =1 \},
%\!\roaacttrue \!=\! \{\xini \!\in \Rn\, | \,\mathbb{P}[\!\lim\limits_{\,\,t \rightarrow \infty} \!d(\psi(t, \xini (\ranvar)),\invset)\! = 0]\! =1 \, \text{a.s.}\},\!
%\end{equation}
\begin{align}
\nonumber
%\roa = \{x_\text{ini} \in \Rn | \,\,\mathbb{P}[\lim\limits_{\,\,t \rightarrow \infty} \text{dist}(\psi(t, \xini),\invset) \rightarrow 0] = 1	\,\,\, \text{a.s.}\}
\!\!\!\!\roaacttrue = &\{\xini \in \Rn | \\[2pt]\label{eq:stochroaset}
%& ||\mathbb{E}[|\psi(t, \spar(\ranvar), \xini (\ranvar))|^{\P}] - \mathbb{E}[|\invset|^{\P}]||_2 \xrightarrow[\,\,t \rightarrow \infty] {} 0\}
&\,\mathbb{P}[\!\lim\limits_{\,\,t \rightarrow \infty} \!d(\psi(t, \xini (\ranvar),\ranvar),\invset)\! = 0]\! =1 \, \text{a.s.}\}
\end{align}
where $\mathbb{P}$ denotes probability, a$.$s$.$ stands for almost surely in $\ranvar$ and $d$ is the distance measured in a chosen norm (e$.$g$.$ the Euclidean norm).
%,

%%%%%%%%%%%%%%%%%%%%%%%%%%%%%%%%%%%%%%%%%%%%%%%%%%%%%%%%%%%%%%%%%%%%%%%%%%%%

\subsection{Polynomial Chaos Expansion}
\label{sec:pce}
%\subsection{Background on PCE}

Polynomial Chaos (PC) expansion can be used to approximate stochastic processes with finite second moment (which includes most stochastic processes of the physical world \cite{Xiu2003}) by a higher dimensional set of deterministic equations. Most of the notations and definitions used in this section can be found e$.$g$.$ in \citeasnoun{Sullivan2015}, \citeasnoun{Lemaitre2010}.  

The PC expansion is performed within an orthogonal polynomial basis where the basis is chosen according to the type of probability distribution of the random variable in order to obtain optimal (in the $\ltwo$-sense) convergence of the expansion. This is the case if the weighting function of the orthogonality relationship of the polynomial basis is identical to the probability function of a random distribution. Table \ref{tab:askeyscheme} shows some of the orthogonal polynomials and their associated probability distributions.

\begin{table}[H]
	\centering
	\caption{Orthogonal polynomial bases and their associated probability distributions.}
	\vspace{0.08cm}
	\begin{tabular}{p{3cm}p{4cm}}
		\toprule
		Polynomial basis & Probability distribution\\
		\midrule
		Hermite & Gaussian \\
		Legendre & Uniform\\
		Jacobi & Beta\\
		Laguerre & Gamma\\
		Charlier & Poisson\\
		Krawtchouk & Binomial\\
		Hahn & Hypergeometric\\\bottomrule
	\end{tabular}
	\label{tab:askeyscheme}
\end{table}

For a given probability space, an orthogonal polynomial basis is defined as follows.

\begin{defi}
	\label{defi:orthobasis}
	Let $\meas$ be a non-negative measure on $\sspace$. A set of polynomials $\mathcal{Q} = \{\basisg_i | i \in \mathbb{N}\} \subseteq \pring$ is called an orthogonal system of polynomials if for each $i \in \mathbb{N}$, $\pdeg(\basisg_i) = i$, $\basisg_i \in \ltwo(\sspace,\meas)$ and
	\begin{equation}
		\label{eq:normfacgen}
	\langle \basisg_i(\ranvar),\basisg_j(\ranvar)\rangle = \int_{\sspace} \basisg_i(\ranvar) \basisg_j(\ranvar) d\meas(\ranvar)= \normfac_i \delta_{ij},
	%& = \int_{\sspace} \basisg_i(\ranvar) \basisg_j(\ranvar) \fpdf(\ranvar)d\ranvar 
	\end{equation}
	where $\normfac_i  :=  \langle \basisg_i(\ranvar),\basisg_i(\ranvar)\rangle$ are the (non-negative) normalization constants of the basis.
\end{defi}
The orthogonal polynomial basis is constructed using a normalization such that $\basisg_0 = 1$.
For any complete orthogonal basis of the Hilbert space $\ltwo(\sspace,\meas)$ the PC expansion is then defined as follows.

\begin{defi}
	\label{defi:gpce}
	Let $\spartwo(\ranvar) \in \ltwo(\sspace,\meas)$ be a square-integrable vector-valued random variable in $\R^m$, $m \in \mathbb{N}$. The \textit{polynomial chaos expansion} of $\spartwo(\ranvar)$ with respect to the stochastic variable $\ranvar$ is the expansion of $\spartwo(\ranvar)$ in the orthogonal basis $\{\basisg_i\}_{i=0}^{\p}$
	\begin{equation}
	\label{eq:paraexpansion}
	\spartwo(\ranvar) = \sum_{i=0}^{\p} \cspartwo_i \basisg_i (\ranvar) \in \Rn,
	\end{equation}
	with vector valued polynomial chaos coefficients, 
	\begin{equation}
		\bar{y}_i = [\bar{y}_{1_i},...,\bar{y}_{\xdim_i}]^{\trans},
	\end{equation}
	which are obtained from 
	\begin{equation}
	\label{eq:galerkin}
	\cspartwo_i = \frac{\langle \spartwo(\ranvar),\basisg_i(\ranvar) \rangle}{ \normfac_i}.
	\end{equation}
	With $\p \rightarrow \infty$ the series in \eqref{eq:paraexpansion} becomes an exact expansion of $\spartwo(\ranvar)$.
	
\end{defi}
The coefficients $\{\cspartwo_i\}_{i\innz}$ can be obtained by computing the integral in equation \eqref{eq:galerkin} for each component of $\spartwo$ using, e.g$.$, Galerkin projection. 

In the remainder of the paper we will denote any PC expansion coefficient or variable dependent on such with an overbar-notation to distinguish them from the stochastic variables. Moreover, the following notation is used for the coefficients of the PC expansion of $y \in \R^m$.
\begin{align}
\label{eq:notationmean}
\bar{y}_0 &:= [\bar{y}_{1_0},\ldots, \bar{y}_{m_0} ]^{\trans} \in \R^m,\\\label{eq:notationvar}
\bar{y}_J &:= [\bar{y}_{1_1},\ldots,\bar{y}_{m_1},\ldots ,\bar{y}_{1_\p},\ldots, \bar{y}_{m_\p}]^{\trans} \in \R^{m\cdot \p},
\end{align} 
where the elements in $\bar{y}_0$ are called the \textit{mean modes}, and the elements in $\bar{y}_J$ the \textit{variance modes}. Together, they present the \textit{stochastic modes}, denoted by 
\begin{equation}
\label{eq:notationfull}
\bar{y} := \bary. 
\end{equation}

\subsection{PC expansion of stochastic polynomial ODEs}

Applying the PC expansion to \textit{stochastic} dynamical systems results in a \textit{deterministic} representation of the system at the expense of an increased state dimension. More precisely, by expanding the random variables up to order $\p$ and projecting the resulting expansion onto each of the $\p$ basis functions, the $\xdim$-dimensional stochastic system is represented by a $\xdim\cdot(\p+1)$-dimensional deterministic system. 
We use the notation 
\begin{equation}
\label{eq:pcesys}
\cxdot := \pcesys(\cx),
\end{equation}
where $\cx \in \R^{\xdim(\p+1)}$ is the vector of PC expansion coefficients, and $\pcesys: \R^{\xdim(\p+1)} \rightarrow \R^{\xdim(\p+1)}$, to refer to the dynamics resulting from the PC expansion of a stochastic system \eqref{eq:sys}.

The expansion is demonstrated for an example system where $\xdim = 1$.
\begin{equation}
\label{eq:simplesys}
\dot{x}(t,\ranvar) = \spar(\ranvar) x^3(t,\ranvar) .
\end{equation}
Expanding \eqref{eq:simplesys} and dropping the $(\ranvar)$ and $(t)$-notation for clarity results in
\begin{equation}
\label{eq:generaldyn}	
\sum_{i=0}^{\p} \cxdot_i \basisg_i=\sum_{j,k,l,m=0}^{\p} \cspar_j \cx_{k} \cx_{l} \cx_{m}\basisg_j \basisg_{k}\basisg_{l} \basisg_{m} .
%\sum_{i=0}^{\p} \cxdot_i \basisg_i= \sum_{j=0}^{\p} \sum_{k=0}^{\p} \sum_{l=0}^{\p}  \cspar_j \cx_{k} \cx_{l}\basisg_j \basisg_{k}\basisg_{l} + \ldots
\end{equation}
Projecting \eqref{eq:generaldyn} onto the $q$-th basis polynomial we obtain $q$ deterministic differential equations
\begin{equation}
\label{eq:projexample}
\sum_{i=0}^{\p} \cxdot_i \langle\basisg_i,\!\basisg_q\rangle\! = \!\!\! \sum_{j,k,l,m=0}^{\p}\!\!  \cspar_j \cx_{k} \cx_{l} \cx_{m}\langle\basisg_j \basisg_{k}\basisg_{l}\basisg_{m},\!\basisg_q\rangle.
%\sum_{i=0}^{\p} \cxdot_i \langle\basisg_i,\basisg_m\rangle =  \sum_{j=0}^{\p} \sum_{k=0}^{\p} \sum_{l=0}^{\p}  \cspar_j \cx_{k} \cx_{l}\langle\basisg_j \basisg_{k}\basisg_{l},\basisg_m\rangle + \ldots
\end{equation}
This expression motivates the introduction of a tensor notation, where we call 
\begin{equation}
\label{eq:galerkintensor}
\gt_{ij..q} = \frac{\langle \basisg_i \basisg_j\! \cdot\!\cdot \cdot ,\basisg_q \rangle}{ \normfac_q},
\end{equation}
the rank-$\rr$ Galerkin tensor, where $\rr$ is the monomial degree. It is a sparse tensor, and a function of the chosen polynomial basis functions which results in constant entries. Even though its size increases rapidly with increasing polynomial degree and truncation order, computing the tensor is a one-time cost. It can be computed once offline and then stored for dynamic computations.
Using the tensor notation in \eqref{eq:galerkintensor}, equation \eqref{eq:projexample} results in
\begin{equation}
\cxdot_q= \sum_{j,k,l,m=0}^{\p}  \cspar_j \cx_{k} \cx_{l}  \cx_{m} \gt_{jklmq}. 
%\cxdot_m=  \sum_{j=0}^{\p} \sum_{k=0}^{\p} \sum_{l=0}^{\p}  \cspar_j \cx_{k} \cx_{l} \gt_{jklm} +\ldots
\end{equation}

\subsection{PC expansion of moments}

In the PC framework the moments of a random variable or stochastic process can be retrieved from the expansion coefficients.
Let $x(t,\ranvar) \in \Rn$ be a solution trajectory of a vector valued stochastic system.
With the notation in \eqref{eq:paraexpansion} the $\P$-th moment, where $1\leq\P<\infty$, can be obtained from 
%\begin{align}\nonumber
%\mathbb{E}[|x(t,\ranvar) |^{\P}]&= \sum_{d=1}^{\xdim}\sum_{i=0}^{\infty}\sum_{j=0}^{\infty}\cdots\\ \label{eq:highmom}
%&\cdots \sum_{\P=1}^{\infty}\cx_{d_i}\cx_{d_j}\cdots\cx_{d_\P} \langle\basisg_i\basisg_j\cdots\basisg_{\P}\rangle\basvec_d
%\end{align}
%\begin{align}\nonumber
%\mathbb{E}[|x(t,\ranvar) |^{\P}]&= \sum_{i=0}^{\infty}\sum_{j=0}^{\infty}\cdots\\ \label{eq:highmom}
%&\cdots \sum_{\P=1}^{\infty}\cx_{i}\cx_{j}\cdots\cx_{\P} \langle\basisg_i\basisg_j\cdots\basisg_{\P}\rangle
%\end{align}
\begin{align}\nonumber
%\mathbb{E}[|x(t,\ranvar) |^{\P}]\!=\! \sum_{i=0}^{\p}\sum_{j=0}^{\p}\! \cdot\!\cdot\!\cdot\! \sum_{\P=0}^{\p}\cx_{i}\cx_{j}\! \cdot\!\cdot\!\cdot\! \cx_{\P} \langle\basisg_i\basisg_j\! \cdot\!\cdot\cdot ,\basisg_{\P}\rangle
\mathbb{E}[|x(t,\ranvar) |^{\P}]&=\!\!\!\sum_{i,j,..,\P=0}^{\p} \!\!\!\cx_{i}(t)\cx_{j}(t) \cdot\cdot\cdot \cx_{\P}(t) \langle\basisg_i\basisg_j \cdot\cdot\cdot ,\basisg_{\P}\rangle\\[2pt]\label{eq:highmom}
& =: \cmomp(\cx).
\end{align}
In particular, for the first moment, i$.$e$.$ the mean of $x(t,\ranvar)$, equation \eqref{eq:highmom} results in
\begin{equation}
\label{eq:meanderiv}
\mean(x(t,\ranvar)) := \mathbb{E}[x(t,\ranvar)]  = \langle x(t,\ranvar), \basisg_0 \rangle =  \cx_0(t).
\end{equation}
For the variance of $x(t,\ranvar)$ we obtain
%\begin{equation}
%\label{eq:variance}
%\varvar := \mathbb{E}[|x(t,\ranvar) - \mathbb{E}[x(t,\ranvar)] |^2] = \sum_{d=1}^{\xdim}	\sum_{j =1}^{\infty}  \cx_{d_j}^2 \langle \basisg_j^2 \rangle \basvec_d
%\end{equation}  
\begin{equation}
\mathbb{E}[|x(t,\ranvar) - \mathbb{E}[x(t,\ranvar)]|^2] = \sum_{j =1}^{\p}  \cx_{j}^2(t)  \normfac_j, % 
\end{equation}  
and for the covariance matrix $\varvar$ of $x(t,\ranvar)$
\begin{equation}
\label{eq:variance}
\varvar:= \sum_{j =1}^{\p}  \cx_{j}(t)  \cx_{j}^{\trans}(t)  \normfac_j, % 
\end{equation}  
where, in particular, we have for each entry of the matrix 
\begin{equation}
\varvar_{kl} = \sum_{j =1}^{\p} \cx_{k_j}(t) \cx_{l_j}(t)  \normfac_j.
\end{equation}

%\begin{align}\nonumber
%\mathbb{E}[|x(t,\ranvar) |^2]  &= \! \sum_{d=1}^{\xdim}\mathbb{E}[|\! \sum_{\i=0}^{\infty} \cx_{d_i}\!(t) \basisg_i|^2]\basvec_d \\\nonumber
%&=\!\sum_{d=1}^{\xdim} \sum_{j =0}^{\infty}  \cx_{d_j}^2 \langle \basisg_j^2 \rangle \basvec_d\\\label{eq:varderiv}
%&=\!\sum_{d=1}^{\xdim} ( \cx_{d_0}^2 + \sum_{j =1}^{\infty}  \cx_{d_j}^2 \langle \basisg_j^2 \rangle) \basvec_d
%\end{align}

\subsection{Truncation error}

For practical purposes, a PC expansion needs to be truncated for a specified order $\p$. As the expansion series is $\ltwo$-convergent for second order processes, low orders of $\p$ are in general sufficient to keep the error introduced by the truncation small and represent the original system sufficiently well \cite{Sullivan2015,Xiu2002}. More analysis of the effect of the truncation order and investigation of various undesired effects that truncated systems can exhibit can be found in \citeasnoun{Field2004}.

In the remainder of the paper the following working assumption will be made.

\begin{ass}
	\label{ass:accuracy}
	There exists a finite truncation order $\p$ such that the stochastic system \eqref{eq:sys} is accurately represented by its truncated PC expansion \eqref{eq:pcesys}.
	%The PC expanded system \eqref{eq:pcesys} truncated at order $\p$ is a sufficiently accurate representation of the stochastic system \eqref{eq:sys}, in the sense that the truncation does not affect the moment behavior of \eqref{eq:pcesys}.
\end{ass}
In case a guaranteed accuracy of the truncated system is required the truncation error can be upper bounded and added to the expansion as model uncertainty, see, e$.$g$.$, \citeasnoun{Muhlpfordt2018a}, and \citeasnoun{Fagiano2011} for details. In \citeasnoun{Lucia2017} the effects of the truncation error on the stability of the moments of linear stochastic systems are investigated and conditions are proposed to factor the approximation error into a robust controller design.

\section{Stability of Stochastic Systems}

We are  interested in analysing the stability properties of the equilibrium set of a stochastic system \eqref{eq:sys} by means of its PC expansion \eqref{eq:pcesys}. In order to draw conclusions from the stability properties of the PC expansion on the stability of the stochastic system, a connection between the behavior of both systems needs to be established.

\subsection{Relationship of equilibria}
\label{sec:equilibria}
Before stating the notions of stability we first show the relationship between the equilibria of \eqref{eq:sys} and \eqref{eq:pcesys}.

\begin{lemm}
	\label{lem:eqpoint}
	If the PC expanded system has an equilibrium point $\cxep  \in \R^{\xdim(\p+1)}$ then the stochastic system has the equilibrium set given by uncertainty-dependent elements, $\invset = \{x \in \Rn \, | \, x \in \xep(\ranvar) \sim  \probdist(\cmomp(\cxep))\}$.
%	If the PC expanded system has an equilibrium point $\cxep  \in \R^{\xdim(\p+1)}$ then the stochastic system has an equilibrium set of uncertainty-dependent equilibrium point with mean $\cxep{_0}$ and covariance $\varvar = \sum_{j=1}^{\p} \cxep{_j} \cxep{_j}^{\trans} \normfac_j$.
\end{lemm}
\textbf{Proof.} The components $\cxep{_i}, i=0,...,(\p+1)$ of the equilibrium point $\cxep$ represent the random variable $\xep$ by the expansion relation in \eqref{eq:paraexpansion}, such that $\xep(\ranvar) = \sum_{i=0}^{\p} \cxep{_i} \basisg_i(\ranvar)$. The moments of $\xep$ are then given by the $\cxep$ through the relation in equation \eqref{eq:highmom}. By the definition of the equilibrium set in \eqref{eq:invset}, every element $x$ belonging to the distribution of $\xep$ is an element of $\invset$. \hfill $\square$

Due to Lemma \eqref{lem:eqpoint} the task of analysing the stability of the uncertainty-dependent equilibrium point of the stochastic system converts to the well-known problem of analysing the stability of an equilibrium point of a deterministic system. Moreover, it emphasizes the important aspect that an equilibrium point of the PC expanded system not only corresponds to an equilibrium set of the stochastic system but also contains the statistical information of the set.
Note that the location of $\cxep$ can be easily obtained by simulating a trajectory of \eqref{eq:pcesys} with initial state in the region of interest.
%Note that this Lemma allows through the tools of PC expansions to investigate the stability properties of a stochastic system with uncertainty-dependent equilibrium sets using the efficient methods that formerly could only be used for systems with uncertainty-independent equilibria. 

\begin{rema}
	\label{rema:eqpoint}
	If the variance modes of $\cxep$ are zero, i$.$.e$.$ $\cxep{_J} = 0$, then the stochastic system has an uncertainty-independent equilibrium point located at $\xep = \cxep{_0}$. The equilibrium set $\invset$ thus only contains one element. Moreover, if all stochastic modes are zero, $\cxep = 0$, then also $\xep = 0$.
\end{rema}

\begin{rema}
	The condition in Lemma \ref{lem:eqpoint} is only sufficient and the reverse does not hold - if the stochastic system has an equilibrium set there is not necessarily one corresponding equilibrium point of the PC expanded system. This is for example the case when the stochastic system has a limit cycle, in which case also the PC expanded system can have oscillating equilibrium states.
\end{rema}

Based on this relationship between the equilibria we propose a connection between certain stability notions which are specified for each system in the following. 
% We consider the class of stochastic systems with uncertainty-dependent equilibrium points and treat the class of stochastic systems with uncertainty-independent equilibrium points as special cases of the former.
%We consider both the case where the zero solution $x(t) \equiv 0$ is an equilibrium point (EP) of both the nominal and the stochastic system and the general case where the location of the EP is affected by the uncertainty and differs from the EP for the nominal system.
%, and show the connection to almost sure stability for the class of systems \eqref{eq:sys} under consideration. 
%

%\begin{rema}
%	Lemma \ref{lem:eqpoint} implies that if the PC expansion has an equilibrium point then this corresponds to a set of uncertainty-dependent equilibrium points for the stochastic system. The reverse is in general not true, even for the class of second order processes considered here. This can easily be seen by considering 
%\end{rema}

\subsection{$\P$-th moment boundedness and stability}
For stochastic systems there are various concepts of stability ranging from weaker forms such as stability in probability to stronger forms such as $\P$-th moment stability up to almost sure stability, see, e$.$g$.$ \citeasnoun{Kozin1969} for an overview. In the following we focus on $\P$-th moment boundedness and stability of stochastic systems where we employ the definitions as found in, e$.$g$.$, \citeasnoun{Khasminskii2012}, \citeasnoun{Wu2004}, \citeasnoun{Khalil2001}:
\begin{defi}
	\label{def:momentstab}
	The solutions of \eqref{eq:sys} are called ultimately bounded in the $\P$-th moment if there exists a constant $c>0$ such that for any $b>0$ there exists a $T=T(b)>0$ such that
%		\item bounded in the $\P$-th moment if for any $a > 0$ there exists a constant $c = c(a) >0$ such that 
%		\begin{equation}
%			|\xini|<a \,\, \rightarrow \,\, \mathbb{E}[|x(t,\ranvar)|^{\P}]<c \,\, \forall t \geq 0 
%		\end{equation}
		\begin{equation}
		\label{eq:actultbounded}
			|\xini|<b \,\, \rightarrow \,\, \mathbb{E}[|x(t,\ranvar)|^{\P}]<c, \quad \forall t \geq T. 
		\end{equation}
	Further, if there is only one element in $\invset$ then let this element, without loss of generality, be the zero point. This zero point is called 
	\begin{itemize}
		\item stable in the $\P$-th moment, if for each $\epsilon >0$, there exists a $\delta >0$ such that 
		\begin{equation}
		|\xini|<\delta \,\, \rightarrow \,\,  \mathbb{E}[|x(t,\ranvar)|^{\P}]<\epsilon, \quad \forall t \geq 0 ,
		\end{equation}
		\item asymptotically stable in the $\P$-th moment, if it is $\P$-th moment stable and, further,  
		\begin{equation}
		\label{eq:actasympstab}
		|\xini|<\delta \,\, \rightarrow \,\, \mathbb{E}[|x(t,\ranvar)|^{\P}] \rightarrow 0 \,\, \text{as } t \rightarrow \infty.
		\end{equation}
	\end{itemize}
\end{defi}

%$p$-moment asymptotic stability thus expresses the vanishing of all moments up to the $p$-th at the zero point. 
We now define a suitable notion of stability for the PC expanded system. As we are interested in equilibrium points of the PC expansion and, further, the PC expanded system is deterministic, we use stability in the sense of Lyapunov. 
%Equation \eqref{eq:highmom} provides a connection between the moments of the solution of a stochastic system and the PC expansion coefficients of the solution. In the following we exploit this connection for analysing moment stability. To this aim we first define suitable notions of stability for a PC expanded system \eqref{eq:pcesys}.
\begin{defi}
	\label{defi:pcestability}
	The equilibrium point $\cxep$ of \eqref{eq:pcesys} is locally stable if for each $\epsilon >0$ there exists a $\delta>0$ such that
	\begin{equation}
	\label{eq:stabtheo1}
	|\cxini|<\delta \Rightarrow |\cx(t)-\cxep|<\epsilon, \quad \forall t>0.
	\end{equation} 
	Further, $\cxep$ is locally asymptotically stable if it is locally stable and $\delta$ can be chosen such that
	\begin{equation}
	\label{eq:stabtheo2}
	|\cxini|<\delta \Rightarrow |\cx(t)-\cxep|\rightarrow 0 \,\,\, \text{as} \,\,\, t \rightarrow \infty.
	\end{equation}
\end{defi}
With Definition \ref{defi:pcestability} we find the following result for the stochastic system.
\begin{thm}
	\label{theo:pcestability}
	Let the system \eqref{eq:pcesys} with $\pcesys: \bar{D} \rightarrow \bar{D} \subseteq\R^{\xdim\cdot(\p+1)}$ be the PC expansion of the stochastic system \eqref{eq:sys}. If the equilibrium point $\cxep \in \bar{D}$ is locally asymptotically stable then the solutions of the stochastic system \eqref{eq:sys} are ultimately bounded in the $\P$-th moment in a neighborhood of $\invset$. If, further, $\cxep$ represents a $\invset$ containing a single point, then \eqref{eq:sys} is locally asymptotically stable in the $\P$-th moment.
	%the equilibrium point $\cxep$ is the zero point, $\cxep = 0$, then \eqref{eq:sys} is locally asymptotically stable in the $\P$-th moment.
\end{thm}
\textbf{Proof.} If $\cxep$ is an equilibrium point of \eqref{eq:pcesys} then every trajectory $\cx(t)$ in a neighborhood of $\cxep$ will eventually converge to $\cxep$. As all components $\cx_i(t)$ in this case converge to a finite value, so does every term in the expression in \eqref{eq:highmom} and thus $\mathbb{E}[|x(t,\ranvar) |^{\P}]$ will eventually converge to a finite value, which is given by inserting $\cxep$ into the right hand side of equation \eqref{eq:highmom}. The ultimate boundedness of the $\P$-th moment as defined in \eqref{eq:actultbounded} follows. 
If the equilibrium point $\cxep$ represents an $\invset$ consisting of a single point then this implies that $\cxep{_J} =0$ (see Remark \ref{rema:eqpoint}). Thus, every component of $\cx_J(t)$ will converge to zero and every component of $\cx_0(t)$ will converge to $\cxep{_0}$ as $t \rightarrow \infty$. Assuming without loss of generality $\cxep{_0} = 0$, it follows that equation \eqref{eq:highmom} converges to zero and thus equation \eqref{eq:actasympstab} holds. \hfill $\square$
%	Asymptotically stable: If there exists a $\delta>0$ such that \eqref{eq:stabtheo2} is satisfied then, with every solution $\cx_i(t)$ converging to zero as $t\rightarrow \infty$, also every term in the sum in \eqref{eq:highmom} and thus the sum itself converges to zero. It follows that the right-hand side in \eqref{eq:highmom} converges to zero and the condition for asymptotic stability in Definition \ref{def:momentstab} is satisfied.

\begin{rema}
	Note that the reverse is not true: ultimately bounded solutions of the stochastic system \eqref{eq:sys} do not imply a convergence of the components $\cx(t)$ to finite values. One example for this is readily provided by systems with a stable limit cycle. The trajectories in a neighborhood of the limit cycle converge to the limit cycle and thus are locally ultimately bounded, however the PC expansion coefficients $\cx_i(t)$ do not converge to an equilibrium point but instead remain ultimately bounded to a set as well.
	%One might also be able to come up with an example in which the the solutions of \eqref{eq:sys} are ultimately bounded and not by a limit cycle, however the solutions of its PC expansion converge to a limit cycle in the variance modes.
\end{rema}

Theorem \ref{theo:pcestability} allows us to obtain information about the behavior of the stochastic system by analysing the local stability properties of an equilibrium point $\cxep$ of the PC expanded system. 
In the following we formulate the criteria with which the attractive region of $\cxep$ can be obtained.

\section{PC Expansion-based Region of Attraction Analysis}
\label{sec:verialgos}

In this section we first define the ROA of an equilibrium point $\cxep$ of the PC expanded system and state the criteria with which an inner estimate of it can be obtained. We then show how this ROA translates to an inner estimate of $\roaacttrue$, the ROA of the stochastic system. 
Finally, optimization programs to maximize inner estimates of both ROAs are proposed.
%Implementations in form of optimization program to maximize an inner estimate of the ROA. We then propose a second optimization program with which an estimate of the ROA of the original stochastic systems in terms of its statistical properties can be obtained. 

\subsection{Formulation of the ROA based on a PC Expansion}

Let the ROA of $\cxep$ be defined by the set
\begin{equation}
	\label{eq:pceroa}
	\roatrue\!= \{\cxini\! \in \R^{\xdim\cdot(\p+1)} | \lim\limits_{\,t \rightarrow \infty}\!d(\bar{\flow}(t,\cxini),\cxep)\! = 0\},\!
\end{equation}
where $\bar{\flow}(t,\cxini)$ denotes the solution of the PC expanded system at time $t$ with initial state $\cxini$.
An inner estimate of $\roatrue$, denoted by $\roa$, is then obtained from the following arguments.
\begin{thm}
	\label{thm:varinball}
	Let $\bar{D} \subset \R^{\xdim\cdot(\p+1)}$ be a compact domain containing $\cxep$ and let $\lyp$ be a continuously differentiable function $\lyp(\cx): \bar{D} \rightarrow \R$. For a scalar $\vsls >0$ let $\slsvr = \{ \cx \in \bar{D} \, | \, \lyp(\cx)\leq \vsls \}$ be the $\vsls$-sublevel set of $\lyp$. If $\lyp$ satisfies
	\begin{align}\label{eq:lypcond1}
	&\lyp(\cx) > 0 \quad \forall \cx \in \slsvr\backslash \{\cxep\}, \,\quad \lyp(\cxep) = 0,\\\label{eq:lypcond2}
	&\dlyp(\cx) < 0 \quad \forall \cx \in \slsvr\backslash \{\cxep\},
	\end{align}
	then $\lyp$ is a Lyapunov function and every trajectory $\cxini$ starting in $\slsvr$ will converge to $\cxep$ as $t\rightarrow \infty$. Thus, the set $\roa = \{\cxini \in \bar{D} | \cxini = \cx,\,\, \forall \cx \in \slsvr\}$  is an inner estimate of $\roatrue$.
\end{thm}
The proof uses Lyapunov arguments which are standard in ROA analysis and can be found, e$.$g$.$ in \citeasnoun{Khalil2001}, and \citeasnoun{Tan2006a}. The novelty here is the application, for the first time to the best of the authors' knowledge, to the PC expanded system and, as shown in the following, the connection of $\roa$ to the stability properties of the original stochastic system.

%\begin{lemm}
%	Let $\vsls >0$ be a scalar and let $\slsvr = \{ \cx \in \bar{D} \, | \, \lyp(\cx)\leq \vsls \}$ be a sublevel set of a function $\lyp$. If $\lyp$ satisfies the conditions of Theorem \ref{thm:varinball} and $\slsvr \subseteq \bar{D}$
%	then every trajectory $\cxini$ starting in $\slsvr$ will converge to $\cxep$ as $t\rightarrow \infty$. Thus, $\slsvr$ is an inner estimate $\roa$ of $\roatrue$, the region of attraction of $\cxep$. 
%\end{lemm}

%\begin{proof}
%	By satisfying the conditions stated in Theorem \ref{thm:varinball} any trajectory starting in the set $\slsvr\backslash \slsfg$ will eventually converge to and enter the set $\slsfg$. As $\slsfg$ is positively invariant the trajectory will remain inside the set. \textcolor{red}{is this needed?} 
%\end{proof}

 Theorem \ref{thm:varinball} presents a criterion for a set $\roa$ to be an estimate of the ROA, where $\roa$ is in terms of the PC expansion coefficients. We now provide the means to infer information about $\roaacttrue$, the ROA of the equilibrium set $\invset$ of the stochastic system, from $\roa$. More precisely, we show how the inner estimate $\roa$ translates into an inner estimate $\roaact$ of the stochastic ROA. Recalling the expression \eqref{eq:stochroaset} for the ROA of the equilibrium set $\invset$ of a stochastic system,  
 %and leveraging the relation \eqref{eq:highmom} of a random variable and its PC expansion coefficients in the computation of the moments,
 the following arguments can be made.
%As the PC expansion coefficients provide information on the moments, in particular on mean and variance, we obtain the following result for the ROA of $\invset$.

\begin{lemm}
	\label{prop:roaset}
	Let $\roa$ be an inner estimate of the ROA of $\cxep$, $\roa \subseteq \roatrue$ . Then the set 
	\begin{equation}
	\label{eq:roaact}
	\!	\roaact\! =\! \{\xini\! \in \Rn | \xini(\ranvar)\! \sim\! \probdist(\cmomp(\cxini)\!),\forall \cxini\! \in \roa\},\!\!\!
	\end{equation} 
	is a subset of the ROA of $\xep$, $\roaact \subseteq \roaacttrue$.
	%\mathbb{E}[x] = x_0, \mathbb{E}[x - \mathbb{E}[x]]^2 = \sum_{j =1}^{\infty}  \cx_{j}  \cx_{j}^{\trans}  \normfac_j, [\cx_0,\cx_J] \in \roa\}$ is an estimate of $\roaact$.
	%	consisting of all elements from the distributions with mean $\cx_0 \subseteq \roa$ and covariance $\varvar \subseteq \roa$ is an estimate of $\roaact$.
	%\{x \in \Rn \, | \, x = \sum_{i=0}^{\p} \cx_i \basisg_i, \, \forall \cx_i \in \roa \}$
\end{lemm}

\textbf{Proof.} We first establish the relationship between $\xini(\ranvar)$ and $\cxini \in \roa$.
The PC coefficients $\cxini \in \roa$ represent the stochastic variables $\xini(\ranvar)$ by the relation \eqref{eq:paraexpansion}, such that any $\xini^{\sharp}(\ranvar) \in \roaact$ is given by $\xini^{\sharp}(\ranvar) = \sum_{i=0}^{\p} \cxini^{\sharp}{_i}\basisg_i(\ranvar)$. For this $\xini^{\sharp}(\ranvar)$, from equation \eqref{eq:highmom} the moments are given by $\momp(\xini^{\sharp})=\cmomp(\cxini^{\sharp})$. This reasoning holds for all $\xini \in \roaact$.

We now turn to prove $\roaact \subseteq \roaacttrue$. 
Recall, that from Theorem \ref{thm:varinball} we have $\cxini^{}\in\roa \Longrightarrow \lim\limits_{\,\,t \rightarrow \infty}\bar{\flow}(t,\cxini)= \cxep$. Let further $\cx(t) = \bar{\flow}(t,\cxini)$ and $x(t,\ranvar) = \flow(t,\xini(\ranvar),\ranvar)$.
With equation \eqref{eq:highmom} and the results from Theorem \ref{theo:pcestability}, it follows that if $\cxini \in \roa$ then
\begin{align}
\nonumber
&\mathbb{E}[|x^{}(t,\ranvar) |^{\P}] =\!\!\!\sum_{i,..,\P=0}^{\p} \!\!\!\cx^{}_{i}(t)\cdot\cdot\cdot \cx^{}_{\P}(t) \langle\basisg_i\cdot\cdot\cdot ,\basisg_{\P}\rangle,\\\nonumber
&\text{and so}\\\nonumber
&\lim\limits_{\,\,t \rightarrow \infty} \sum_{i,..,\P=0}^{\p} \!\!\!\cxep{_i} \cdot\cdot\cdot \cxep{_\P}\langle\basisg_i\cdot\cdot\cdot ,\basisg_{\P}\rangle\\\label{eq:proofeq0}
%,\basisg_{\P}\rangle \\[2pt]
&\quad\quad\quad\quad\quad\quad\quad\quad\quad\quad\quad\quad= \mathbb{E}[|\xep(\ranvar) |^{\P}], 
\end{align}
where $1\leq\P<\infty$ and for a given $\xep(\ranvar)$ and $\P$ the term $\mathbb{E}[|\xep(\ranvar)|^{\P}] $ is a constant.

So far, we have shown the moment convergence of a random variable $\xini(t,\ranvar) \in \roaact$. It remains to show that from this follows $\lim\limits_{\,\,t \rightarrow \infty} \mathbb{P}[d(\flow(t,\xini^{}(\ranvar),\ranvar),\invset) =0]=1$ a$.$s$.$ .

To this end, assume there is a subset $\sspace^{\dagger}\subset\sspace$ for which $\ranvar^{\dagger} \in \sspace^{\dagger}: d(x(t,\ranvar^{\dagger}),\invset)\not\rightarrow 0$ as $t\rightarrow\infty$.
Consider first the case where $x(t,\ranvar^{\dagger}) \rightarrow \infty$ as $t\rightarrow\infty$.
Then 
\begin{align}
\nonumber
&\!\!\mathbb{E}[|x^{}(t,\ranvar) |^{\P}] = \int_{\sspace} |x(t,\ranvar)|^{\P} d\meas(\ranvar)\\ \label{eq:pp1}
&\!\!= \int_{\sspace^{\dagger}} \!\!|x(t,\ranvar^{\dagger})|^{\P} d\meas(\ranvar^{\dagger})\!+\!\int_{\sspace^{\dagger C}}\! |x(t,\ranvar^{\dagger C})|^{\P} d\meas(\ranvar^{\dagger C}),
\end{align}
where $\ranvar^{\dagger C} \in \sspace^{\dagger C}$ and $\sspace^{\dagger C}$ denotes the complement of $\sspace^{\dagger}$, such that $\sspace^{\dagger C}\cup\sspace^{\dagger} = \sspace$.
The first term in equation \eqref{eq:pp1} and by that the $\P$-th moment of $x(t,\ranvar)$ will, however, tend to infinity as t goes to infinity, unless the elements in $\sspace^{\dagger}$ have $\meas$-measure zero.
Consider now the case where $d(x(t,\ranvar^{\dagger}),\invset) \rightarrow c$ as $t\rightarrow\infty$, where $0<c<\infty$ is a constant.
In order to not contradict \eqref{eq:proofeq0} with the expression in \eqref{eq:pp1} we find that either $x(t,\ranvar^{\dagger}) = x(t,\ranvar)$ for all $\ranvar^\dagger = \ranvar$, but this implies $d(x(t,\ranvar^{\dagger}),\invset)\rightarrow 0$ as $t\rightarrow \infty$, or $\mu(\ranvar^{\dagger}) = 0$.
Hence, from moment convergence follows the almost sure convergence of $x(t,\ranvar)$ to $\invset$, such that $\lim\limits_{\,\,t \rightarrow \infty} \mathbb{P}[d(\flow(t,\xini^{\dagger}(\ranvar),\ranvar),\invset) =0]=1$ a$.$s$.$ for all $\xini \in \roaact$ and thus $\roaact\subseteq\roaacttrue$. \hfill $\square$

\subsection{Algorithmic computation of $\roa$}
In the following we present optimization algorithms by which $\roa$ can be computed. 
In order to make the following implementations generalizable, a coordinate shift is introduced, similar to the one proposed in \citeasnoun{Iannelli2018}. The shift is
\begin{equation}
\label{eq:coorshift}
\cz = \cx -\cxep,
\end{equation} 
and it is such that the analysed system is centered around the zero point. Note that while in \citeasnoun{Iannelli2018} $\cxep$ is not known because it depends on the uncertainty, in this formulation $\cxep$ is deterministic and obtained by simulation of the PC expanded system, as mentioned in Section \ref{sec:equilibria}.

Using polynomial functions for $\lyp$, the conditions on the set $\roa$ as stated in Theorem \ref{thm:varinball} are in polynomial form. This allows to employ an approach introduced in \citeasnoun{Parrilo2000}, and formulate the ROA conditions as semi-algebraic set emptiness conditions. These can be efficiently solved through a relaxation to sum-of-squares (SOS) programs employing Stengle's Positivstellensatz \cite{Stengle1974}. Details on the procedure of formulating conditions such as those in Theorem \ref{thm:varinball} and Lemma \ref{lem:eqpoint} as set emptiness conditions and casting them as SOS constraints are omitted for brevity and can be found in, e$.$g$.$ \citeasnoun{Parrilo2000}, \citeasnoun{JarvisWloszek2005}, and \citeasnoun{Topcu2008}. The resulting SOS program consists of polynomial objectives and polynomial constraints. Each of the constraints is a requirement that the polynomial is SOS. The SOS program can then be reformulated as a semidefinite program (SDP) that is convex in the coefficients of the polynomials.

%Following an approach introduced in \cite{Parrilo2000} we formulate the stability conditions as set emptiness questions which are relaxed to sum-of-squares (SOS) programs employing Stengle's Positivstellensatz \cite{Stengle1974}. Similar procedures were previously used for verification of ROA in, among others, \cite{JarvisWloszek2005,Topcu2008}.
Applying the procedure to the conditions on $\roa$ as stated in Theorem \ref{thm:varinball} results in the following SOS optimization program.
% where for clarity we omit the notation of the $\cz$-dependence.
\begin{subequations}
	\label{eq:algoroaepmain}
	\begin{alignat}{3}
	\label{eq:algoroaep}
	& \max_{\lyp(\cz),\mb(\cz),\vsls} \quad \quad \text{vol}(\roa(\cz)) \quad \quad \\[1.2pt]\nonumber
	&\text{subject to }\quad \quad\quad \quad  \quad\,&& \\\label{eq:roacon0}
	&\quad\quad \quad \quad \quad \quad \quad \quad \quad \, \lyp(\cz)  -  \la(\cz)\,\,\,&& \in \sos[\cz],\\\label{eq:roacon1}
	&- \dlyp(\cz)   -  \mb(\cz)(\vsls - \lyp(\cz))  -  \lb(\cz)\,\,\,&& \in \sos[\cz],\\\label{eq:roacon2}
	&\text{}\quad\quad\quad \quad \quad\quad \quad \quad \quad\quad \quad \quad\vphantom{\dlyp} \mb(\cz)\,\,\,&&  \in \sos[\cz],
	\end{alignat}
\end{subequations}
where the multiplier $\mb$ is an SOS polynomial of potentially arbitrarily high degree which results directly from the Positivstellensatz and, once obtained, certifies that the solution of the program adheres to the constraints. 
The term $\la(\cz)$ is an even polynomial with small fixed coefficients (e$.$g$.$, $\la(\cz) = 10^{-4} \cz^{\trans}\cz$), which results from the definiteness of the conditions in \eqref{eq:lypcond1} and \eqref{eq:lypcond2} for all $\cx$ except for $\cxep$. 

In order for the optimization problem \eqref{eq:algoroaepmain} to be convex in the decision variables and thus solvable as an SDP, the following steps are taken.
The set $\slsvr$ is formulated as the sublevel set $\slsvriso =\{\cz \, | \, \lyp (\cz) :=  v(\cz)^{\trans} \Qlyp v(\cz)  \leq 1, \,\, \Qlyp>0\}$ where $v(\cz)$ is the vector of monomials in $\cz$ and $\vsls$ is fixed to 1 as optimizing over $\rho$ is redundant when optimizing over $\Qlyp$. Furthermore, the objective in \eqref{eq:algoroaep} is a generic expression for the volume of the ROA and needs to be replaced by a convex expression. It has been previously observed \cite{JarvisWloszek2005,Tan2006a,Ahbe2018} that higher degree functions $\lyp$ have the potential to verify larger estimates of the ROA. For $\pdeg(\lyp)>2$ the volume of a sublevel set cannot be computed from a convex expression and thus a surrogate that is a computationally tractable measure for the ROA is employed. We use a convex measure in the form of the geometric mean of the eigenvalues of the matrix $\ballmat$ of the sublevel set $\ballset=\{\cz |\, \insideball := \cz^{\trans}\ballmat\cz \leq 1\}$ of a quadratic function $\insideball(\cz)$. The geometric mean of the eigenvalues of a matrix is a monotone function of the determinant, which itself is inversely proportional to the volume of the set. Minimizing the geometric mean of the eigenvalues thus maximizes the volume of a quadratic set. With the constraint that the surrogate set $\ballset$ lies inside the sublevel set $\slsvro$, $\ballset \subseteq \slsvro$, a maximization of the set $\ballset$ leads to the estimate of $\roa$ being increased simultaneously. See \citeasnoun{Ahbe2018} for more details and comparisons of convex measures for the ROA. 
Utilization of this surrogate set requires adding the following constraints to the optimization program~ \eqref{eq:algoroaepmain}:
\begin{subequations}
	\label{eq:surrogatecon}
	\begin{alignat}{3}
	\label{eq:surrogatecon1}
	&-  \mc(\cz) (1 - \insideball(\cz) )  - (1 - \lyp(\cz)) \,\,\,&& \in \sos[\cz],  &&\\\label{eq:surrogatecon2}
	&\text{}\quad \quad\quad\quad\quad\quad\quad\quad\quad\quad\,\,\quad\vphantom{\dlyp}\mc(\cz)\,\,\,&&  \in \sos[\cz].
	\end{alignat}
\end{subequations}
The objective function \eqref{eq:algoroaep} is then replaced by the geometric mean of the eigenvalues of $\ballmat$,
\begin{equation}
\label{eq:newobjfun}
\min_{\lyp,\mb,\mc,\ballmat}\quad \text{det}(\ballmat)^{1/\xdim(\p+1)}
\end{equation}
The resulting optimization program then consists in
\begin{subequations}
	\label{eq:finalprog}
	\begin{alignat}{3}
	&\text{solve } &&\quad \eqref{eq:newobjfun}\\
	&\text{subject to }\vphantom{\dlyp} &&\quad \eqref{eq:roacon0},\eqref{eq:roacon1}, \eqref{eq:roacon2}, \eqref{eq:surrogatecon1},\eqref{eq:surrogatecon2}.
	\end{alignat}
\end{subequations}
This SOS program is bilinear in the multipliers $\mb,$ respectively $\mc$ and $\lyp$, respectively $\ballmat$, which prevents its direct solution as an SDP. However, it can be solved iteratively as an SDP by fixing one of the two bilinearly appearing variables and optimizing over the other, and vice versa. This requires an initial estimate for one of the two variables. Here, an inital estimate for $\slsvro$ is obtained by linearizing the system \eqref{eq:pcesys}, shifted in the coordinates as in \eqref{eq:coorshift}, around the equilibrium point $\cz = 0$ and solving the Lyapunov inequality for the linearized state matrix. The resulting Lyapunov matrix is then suitably scaled (e$.$g$.$ by bisection) to obtain a feasible initial Lyapunov function for the nonlinear system. Similarly, the initial estimate of the matrix $\ballmat$ can be found by using a unit diagonal matrix with a suitable scaling. 
%The existence of feasible initial $\lyp$ for a stable EP for a smooth system is given by Lyapunov's Indirect method an

%Note that the SOS multiplier used to certify the constraints as stated by the Positivstellensatz are not limited in their degree from the theoretical point of view. For practical and computational reasons, however, the degrees need to be fixed which renders the constraints in the program \eqref{eq:algoroaepmain}, \eqref{eq:surrogatecon} sufficient but not necessary for certifying the $\roa$. 

\subsection{Recovering $\roaact$ from $\roa$}

We propose an approach in form of an optimization problem in which the set $\roaact$, as given by Lemma \ref{prop:roaset} for initial conditions with specified stochastic properties, can be recovered from the set $\roa$. 
In particular, the program shows how to obtain a maximized estimate $\roaact$ of the true ROA $\roaacttrue$ from a given set $\roa$. The set $\roaact$ is given by \textit{stochastic} variables $x$, whose statistical properties are given by all possible states of the PC coefficients contained in $\roa$. In the set $\roa$, the mean modes $\cx_0$ and the variance modes $\cx_J$ can be traded off, allowing for a wide range of distributions of $x$ being represented by $\roa$.
In order to obtain a set $\roaact$ of the stochastic system in the $x$ variables, one of the two statistical properties, either the mean or the covariance, of the initial states can be fixed and the set $\roaact$ obtained in terms of the other. We here choose to fix the covariance of the initial states $x$ to a specified level, which is denoted by $\varfix$, and compute $\roaact$ in terms of the mean of $x$. The $\roaact$ obtained in this way will be denoted by $\roazero$ in the following. Since $\mean(x) = \cx_0$ (equation \eqref{eq:meanderiv}), the set $\roazero$ is given by
\begin{equation}
	\roazero = \{ \cx_0 \in \roa\,\, |\,\,\cx \in \roa, \, \,\sum_{j = 1}^{\p}  \cx_{j}  \cx_{j}^{\trans}  \normfac_j = \varfix \}.
\end{equation}
Note that as defined in \eqref{eq:notationfull}, $\cx = [\cx_0,\cx_J]^{\trans}$.

The set $\roazero$ can be computed from a given $\roa$ by the following optimization problem. Let $\roazero$ hereby be represented by the 1-sublevel set of the polynomial function $\roazero:=\{\cx_0\,|\,v(\cx_0)^{\trans} \roamat \, v(\cx_0) \leq 1\}$. The aim is to maximize $\roazero$ inside $\roa$ while keeping the size of the polynomials in \eqref{eq:variance}, representing the covariance of the initial states $\xini$, fixed. 
\begin{subequations}
	\label{eq:actualroaalgo}
	\begin{alignat}{2}
	&\max_{\roamat}\quad \quad\quad\quad \text{vol}(\roazero)  \\[1.5pt]
	&\text{subject to } \\\label{roaalgocon1}
	&\quad \quad\quad\quad v(\cx)^{\trans}\Qlyp\,v(\cx)&&\leq 1,  \\\label{roaalgocon2}
	&  \quad\quad\quad \quad  \quad \,\sum_{j = 1}^{\p}  \cx_{j}  \cx_{j}^{\trans}  \normfac_j &&= \varfix,\\\label{roaalgocon3}
	&  \quad\,\,\, \quad \quad v(\cx_0)^{\trans} \roamat  v(\cx_0) &&\leq 1,\\\label{roaalgocon4}
	& \quad \quad\quad\quad \quad\quad \quad \quad\vphantom{\dlyp} \,\,\,\,  \roamat &&>0,\\
	& \quad \quad\quad\quad \quad\quad \quad \quad\vphantom{\dlyp} \,\,\,\,  \roazero&&\subseteq \slsvr,
	\end{alignat}
\end{subequations}
where $\Qlyp$ is the optimizer of \eqref{eq:algoroaepmain}.
Note that \eqref{roaalgocon2} is a matrix equality constraint with polynomial entries. Since $\varfix$ is a symmetric matrix, equation \eqref{roaalgocon2} results in $\frac{\xdim(\xdim+1)}{2}$ scalar constraints.
As  $\pdeg(\roazero) = \pdeg(\lyp)$, a convex surrogate set similar to that in  \eqref{eq:surrogatecon} is introduced to tractably maximize $\roazero$ for $\pdeg(\roazero) >2$ . To this end we use a quadratic sublevel set in terms of the mean modes, $\ballset_0=\{\cx |\, \cx_0^{\trans}\ballmat_0\cx_0 \leq 1\}$, constrained to remain within $\roazero$. The following constraints are added to program \eqref{eq:actualroaalgo} to give a convex optimization of a lower bound on the volume of $\roazero$.
\begin{subequations}
	\label{eq:ballconstraints}
	\begin{alignat}{1}
	\cx_0^{\trans}\ballmat_0\cx_0&\leq 1,\\
	\vphantom{\dlyp}\ballmat_0 &>0,\\
	\vphantom{\dlyp}\ballset_0 &\subseteq \roazero.
	\end{alignat}
\end{subequations}
The following optimization program shows the implementation of the problem in \eqref{eq:actualroaalgo}-\eqref{eq:ballconstraints} that efficiently obtains an estimate of $\roazero$. Note that the objective function is the volume of the surrogate set, $\ballset_0$.
%\begin{subequations}
%	\label{eq:roazeromax}
%	\begin{alignat}{3}
%	&\max_{s_0,s_{11}\ldots,s_{\xdim\xdim},s_m,\roamat,\ballmat_0} \quad \quad \text{det}(\ballmat_0)^{1/\xdim} \\\nonumber
%	&\text{subject to:}\\\nonumber
%	&-\! s_{0}(\cx)(1 - v(\cx_0)^{\trans}\! \roamat v(\cx_0))  - s_{11}(\cx) (\varfix_{11} - \cx_{1_J}^{\trans}&\Gamma\cx_{1_J})\! -...\\\label{eq:roazeromax2}
%	&...\! - s_{\xdim\xdim}(\cx)(\varfix_{nn} - \cx_{\xdim_J}^{\trans}\Gamma\cx_{\xdim_J})  + (1-v(\cx)^{\trans}\!\Qlyp v(\cx))\! &\in \sosc\\\label{eq:roazeromax3}
%	&\,\,\,-(1-\cx_0^{\trans}\ballmat_0 \cx_0)s_{m}(\cx_0) +(1-v(\cx_0)^{\trans} \roamat \, v(\cx_0))\!  &\in \sosc\\\label{eq:roazeromax4}
%	&\quad \quad\quad\,\, \quad \quad\quad\quad\quad \quad s_0(\cx),s_{11}(\cx)\ldots,s_{\xdim\xdim}(\cx)\!&\in \sosc\\\label{eq:roazeromax5}
%	&\quad\quad\quad\quad\,\,\,\,\quad\quad\quad\quad\quad\quad \quad \quad \quad \quad \quad s_{m}(\cx_0) \!&\in \sos[\cx_0]
%	\end{alignat}
%\end{subequations}
\begin{subequations}
	\label{eq:roazeromax}
	\begin{alignat}{2}
	&\max_{s_0,h_{lk}\ldots,s_m,\roamat,\ballmat_0} \quad \quad \text{det}(\ballmat_0)^{1/\xdim} \\\nonumber
	&\text{subject to:}\\\nonumber
	&-s_{1}(\cx)(1\! -\! v(\cx_0)^{\trans}\! \roamat v(\cx_0)) + (1\!-\!v(\cx)^{\trans}&&\Qlyp v(\cx)) +  \\\label{eq:roazeromax2}
	&\quad \quad\,\,\,+\! \sum_{l=1,k\geq l}^{\xdim} h_{lk}(\cx) (\varfix_{lk}\, - \,\cx_{l_J}^{\trans}\Gamma\cx_{k_J})  &&\!\!\!\in \sosc,\\\nonumber
	&-(1-\cx_0^{\trans}\ballmat_0 \cx_0)s_{2}(\cx_0)+&&\\\label{eq:roazeromax3}
	& \vphantom{\dlyp}\quad \quad\quad\quad\quad\quad+(1-v(\cx_0)^{\trans} \roamat \, v(\cx_0))\!  &&\!\!\!\in \sosc,\\\label{eq:roazeromax4}
	&\vphantom{\dlyp}\quad \quad\quad\quad \quad\,\,\, \,\,\quad\quad\quad\quad\quad\quad\,\,\quad\quad s_1(\cx) \!&&\!\!\!\in \sosc,\\\label{eq:roazeromax5}
	&\vphantom{\dlyp}\quad\quad\quad\quad\quad\quad\,\quad\quad \quad \quad\quad\quad \,\, \,\quad s_{2}(\cx_0) \!&&\!\!\!\in \sos[\cx_0].
	\end{alignat}
\end{subequations}
The objective function is now the volume of the surrogate set $\ballset_0$ represented by the geometric mean of the eigenvalues of the matrix $\ballmat_0$. The vector $\cx_{d_J} :=[\cx_{d_1},...,\cx_{d_\p}]^{\trans}$ contains the variance modes of the d-th dimension with $d =1,\ldots,\xdim$ and $\Gamma = \text{diag}[\normfac_1,...,\normfac_{\p}]$. The sum in the second term of \eqref{eq:roazeromax2} represents the scalar equality constraints given by the matrix equality in \eqref{roaalgocon2}. The polynomials $s_1,s_2$ are the SOS-multipliers, resulting from the application of the Positivstellensatz, which certify the inequality constraints. The polynomials $h_{lk}$ are indefinite multipliers certifying the equality constraints. The highest monomial degree in $v(\cx_0)$ is chosen to be equal to the highest monomial degree of $v(\cx)$ in $\lyp(\cx)$. As the constraint \eqref{eq:roazeromax3} involves only the $\cx_0$ coordinates, the associated multiplier $s_2$ contains polynomial terms only in $\cx_0$. The algorithm has bilinear terms in the SOS-multipliers and $\ballmat_0$, respectively $\roamat$. As is the case in the program in \eqref{eq:finalprog}, we solve \eqref{eq:roazeromax} iteratively. 

%Algorithm \ref{alg:actualroamax} presents an outline of the implementation details.
\begin{rema}
	If $\pdeg(\lyp) =2$ then the optimization can be performed to directly minimize $\text{det}(\roamat)^{1/\xdim}$ without using the surrogate set. This removes the constraints \eqref{eq:roazeromax3} and \eqref{eq:roazeromax5} from the algorithm.
\end{rema}

\begin{rema}
	\label{rema:directrzero}
	In the case of $\varfix =0$, i$.$e$.$ the covariance in the initial state is fixed to zero, $\roazero$ can be obtained directly from the computed estimate $\roa$ by setting all terms containing variance modes to zero. In this case there is no need to solve \eqref{eq:roazeromax}.
\end{rema}

\begin{rema}
	The complementary problem of maximizing the allowed covariance in the initial conditions for a fixed mean can be done by inserting the desired fixed matrix $\roamat$ and moving $\varfix$ into the objective. The objective then consists of the convex expression $\text{det}(\varfix)^{1/\xdim}$ and the resulting problem can be solved without the use of a surrogate set and its associated constraints.
	% The set $\{\cx_{d_J} \,| \, \cx_{d_J}^{\trans}\Gamma\cx_{d_J} \leq \varfix_{d}, \forall d \}$ is then obtained by solving the optimization problem as in Algorithm \ref{alg:actualroamax} with the adjusted objective in the function \Call{maximizeROA}{}.
\end{rema}

%\begin{rema}
%	In line 5 in Algorithm \ref{alg:actualroamax} the first diagonal element of the initial instance of $\roamat$ is set to zero as otherwise the set containment of $\ballmat_0$, which due to its structure has no homogeneous term, is not feasible. 
%\end{rema}

\section{Numerical Examples}
We demonstrate the application of the proposed analysis method to an uncertain Van-der-Pol system and to the dynamics investigated in \citeasnoun{Iannelli2018}. Both dynamics are affected by uncertainty in form of a random variable with a uniform distribution. While the first example locally converges to the zero point for all realizations of the random variable, the dynamics of the second example have an uncertainty-dependent attractive equilibrium point. 

We denote in the following a uniform distribution between the boundary values $u$ and $v$ by Unif$(u,v)$. The choice of a uniform distribution is motivated here by the possibility to compare the results to previous studies. However, any other $\ltwo$-distribution can be considered using the methods presented. Considering other distributions only requires the computation of the Galerkin tensor \eqref{eq:galerkintensor} for the associated polynomial basis.

The numerical results were computed with Matlab 2018b, using the open-source toolbox YALMIP \cite{Lofberg2009} to formulate the SOS programs and the commercial solver Mosek to solve the SDPs.

\subsection{Uncertain Van-der-Pol dynamics}

In the first example, we consider
\begin{align}\nonumber
\dot{x}_1 &= -x_2,\\\label{eq:vdpex1}
\dot{x}_2 &= -\spart(\ranvar)  (1 - x_1^2) \, x_2 + x_1,
\end{align}
where $\spart(\ranvar)\sim \text{Unif}(0.7,1.3)$ is a random variable depending on the stochastic germ $\ranvar \sim \text{Unif}(-1,1)$. In order to obtain optimal convergence properties we use the Legendre polynomial basis for the PC expansion of the dynamics which is the basis associated with uniform probability distributions (see Table \ref{tab:askeyscheme}). 
We expand the dynamics \eqref{eq:vdpex1} in the Legendre basis to obtain the PC expansion coefficient dynamics
\begin{align}\nonumber
\cxdot_{1_q} &= \cx_{2_q}, \\\label{eq:vdpex1pce}
%\cxdot_{2_m} &= -\cspart_m + \sum_{i = 0}^3 \sum_{j = 0}^3\sum_{k = 0}^3\sum_{l = 0}^3\cspart_l \cx_{1_j}\cx_{1_k} \cx_{2_l} \gt_{ijklm} + \cx_{1_m}
\cxdot_{2_q} &= -\cspart_q + \sum_{i,j,k,l = 0}^3\cspart_i\, \cx_{1_j}\cx_{1_k} \cx_{2_l} \gt_{ijklq} + \cx_{1_q}.
\end{align}
The dynamics \eqref{eq:vdpex1pce} have an equilibrium point $\cxep = 0$ and thus the equilibrium set $\invset$ consists of the zero point which shows that the system \eqref{eq:vdpex1}  has an uncertainty-independent equilibrium point at $\xep =0$. This equilibrium point is locally stable for $\spart >0$ and for any fixed $\spart>0$ the true region of attraction is given by the unstable limit cycle encircling the equilibrium point which can be obtained by simulation.

\begin{figure}[h]
	\centering
	\includegraphics[width=0.8\columnwidth]{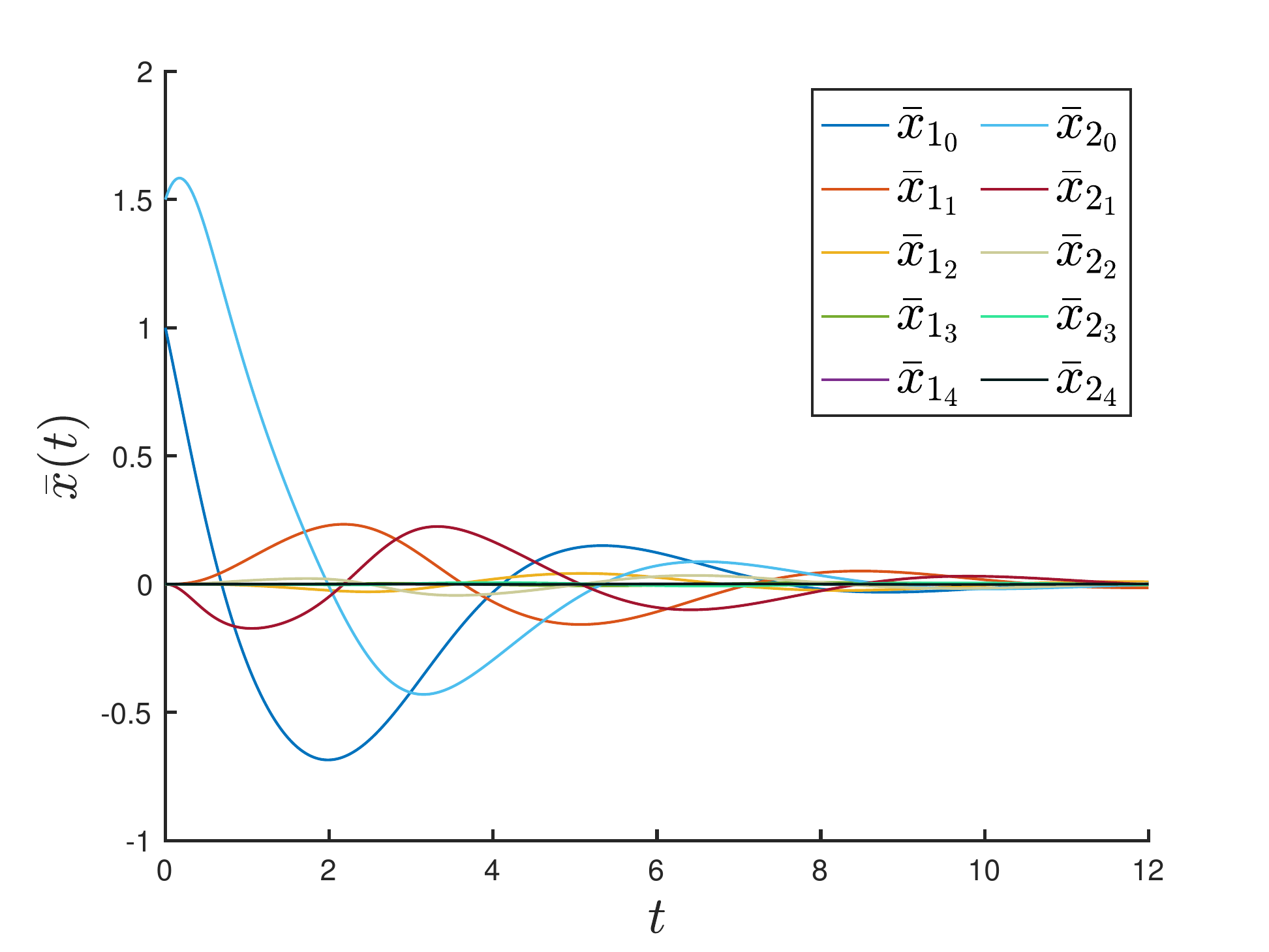}
	\caption{Evolution of the stochastic modes in \eqref{eq:vdpex1pce} starting from the deterministic location $\xini = [1,1.5]^{\trans}$. All modes eventually converge to zero, which is the equilibrium point of the system. }
	\label{fig:vdp1samplesimu}
\end{figure}

\begin{figure}[h]
	\centering
	\includegraphics[width=1\columnwidth]{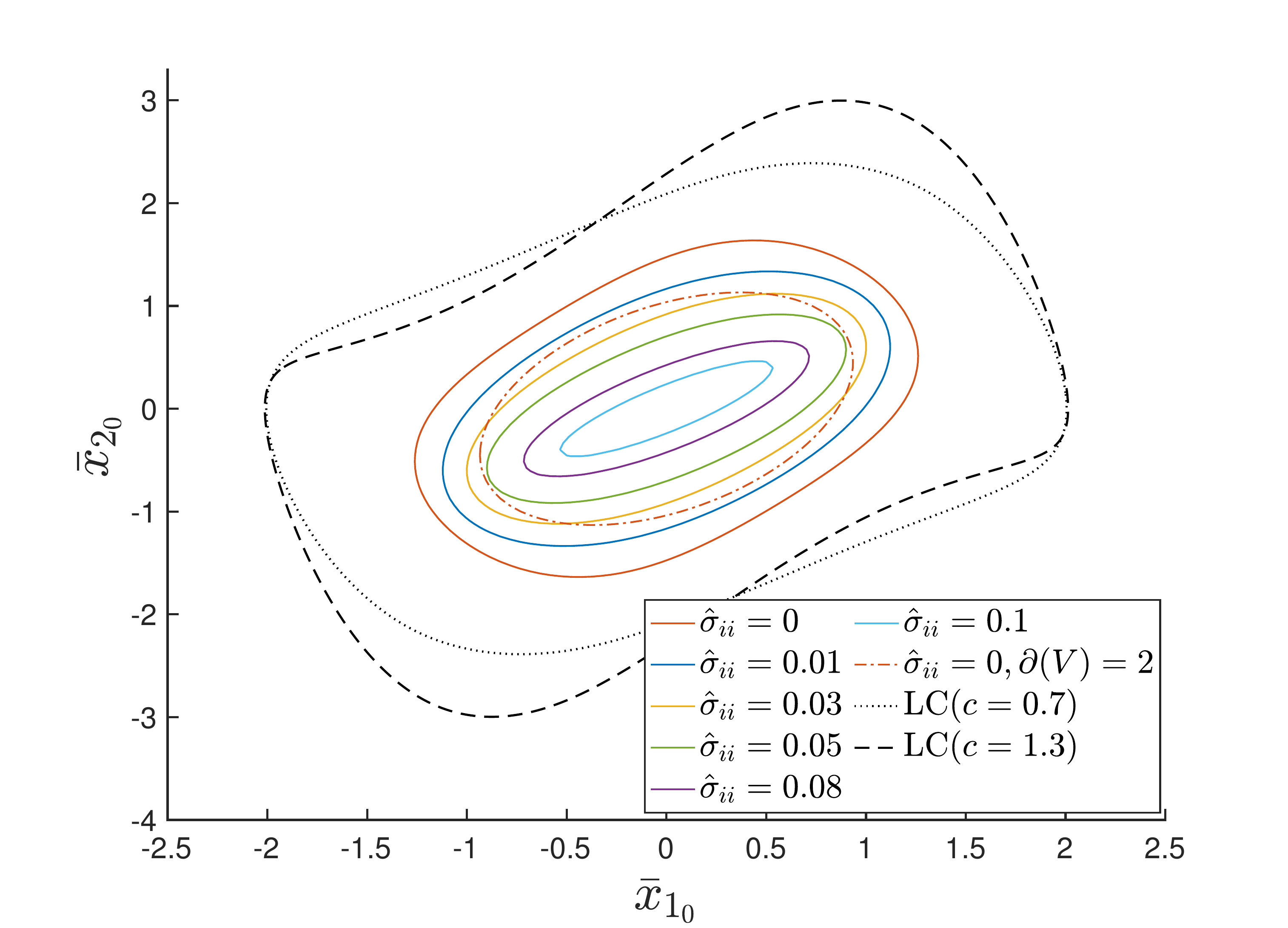}
	\caption{Estimates of $\roazero$ in terms of the initial mean states for various cases of fixed variance on the initial state. The results are obtained from a quartic $\lyp$ for several variance sizes and from a quadratic $\lyp$ for the case of zero minimum variance for comparison. We consider for each case an equal variance in both initial coordinate states. The black dashed and dotted lines show the Van-der-Pol limit cycle trajectory for the extreme values of the uniform distribution of $\spart(\ranvar)$. In this example their intersection provides an upper bound to the true ROA of the system, and thus give an indication on the conservatism associated with the computed inner estimates.}
	\label{fig:vdp1roacases}
\end{figure}

In order to choose the truncation order of the PC expansion, the significance of the first five stochastic modes has been investigated by simulating the dynamics in \eqref{eq:vdpex1pce}. Figure \ref{fig:vdp1samplesimu} shows the evolution of the modes starting from the deterministic initial condition $\xini = [1,1.5]$. As the modes for  $\p>3$ are negligible $\p=3$ has been chosen for the truncation which results in a total of $\p+1 =4$ modes per dimension.
We compute the sublevel set $\roa$ from the optimization program \eqref{eq:finalprog} for $\pdeg(\lyp)=4$, and use the results to compute the ROA estimate $\roazero$ as in \eqref{eq:roazeromax} for different values of fixed variance on the initial condition. We choose a diagonal covariance matrix $\varfix$ with equal diagonal entries. The results are presented in Figure \ref{fig:vdp1roacases} (solid lines). As intuitively expected, the $\roazero$ in terms of the initial state of the mean modes decreases with increasing size of variance in the initial state. Additionally, for comparison of different Lyapunov function degrees, we compute the $\roazero$ estimate with zero initial variance for a quadratic $\lyp$ (red dash-dot line). It can be seen in Figure \ref{fig:vdp1roacases} how the higher degree $\lyp$ returns larger estimates of the ROA in this case. Figure \ref{fig:vdp1roacases} further shows the true ROA of the stochastic system which in this example consists of the intersection set encircled by the two limit cycles resulting from using the extreme realizations of the uncertainty to simulate the system (black dashed and dotted lines).

%\begin{figure}[H]
%	\centering
%	\includegraphics[width=0.8\columnwidth]{figures/vdp_samplesimu.eps}\vspace{-0.5cm}
%	\caption{Sample trajectories starting at the boundary of the $\roa_0$ estimate obtained from \eqref{eq:actualroaalgo} for zero variance in the initial state. For each initial state the trajectories of the real system for the extreme values of the distribution, $\spart \in \{0.7,1.3\}$ and the trajectory of the mean modes of the PC expansion are shown. For the purpose of illustration the trajectory starting from an initial state not inside the full uncertain ROA is presented.  }
%	\label{fig:vdp1sampletrajs}
%\end{figure} 

\subsection{Dynamics with uncertainty dependent equilibria}
\begin{figure}[h]
	\centering
	\includegraphics[width=1\columnwidth]{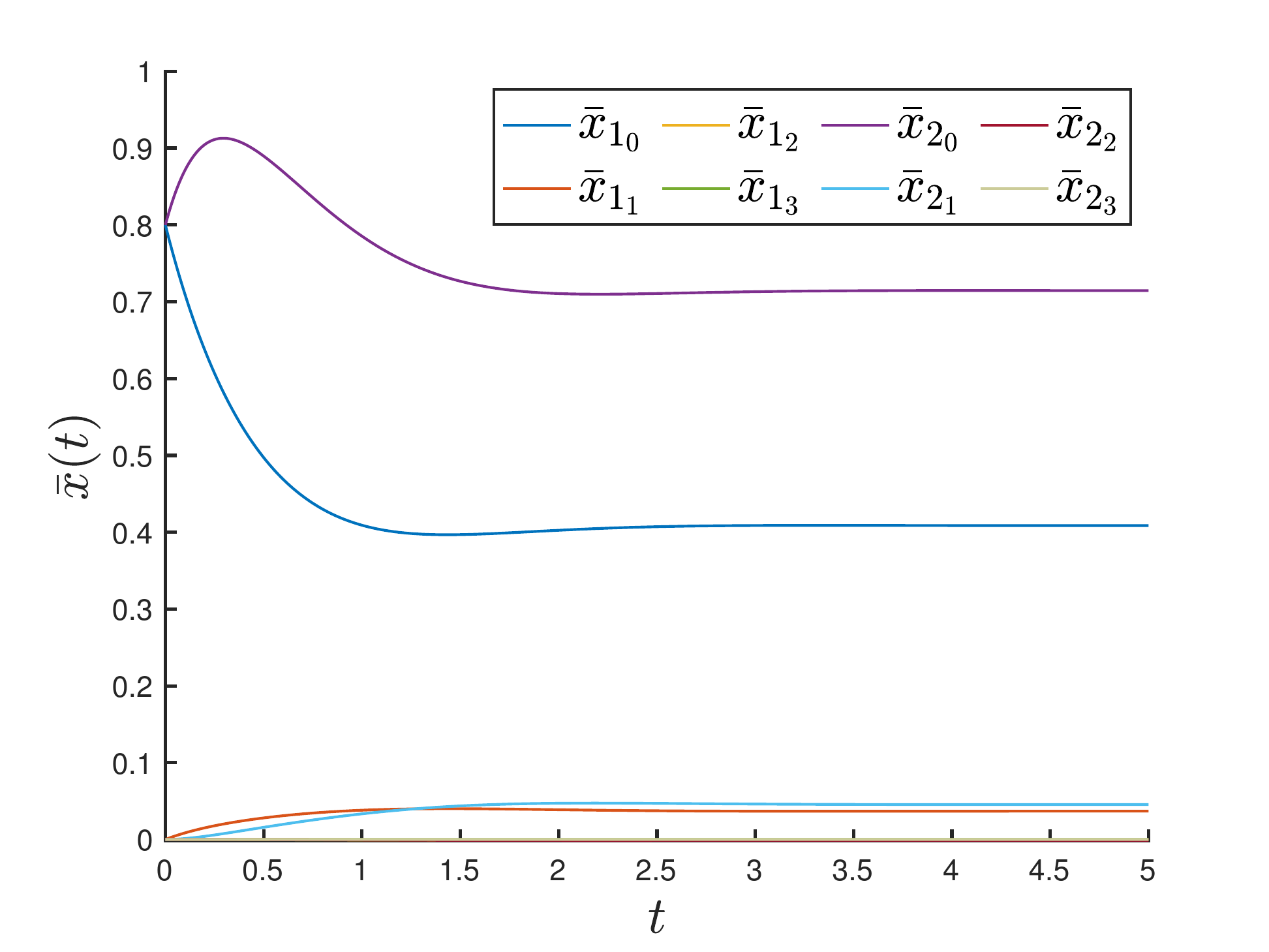}
	\caption{Evolution of the stochastic modes in \eqref{eq:vdpex2} starting from the deterministic location $\xini = [0.8,0.8]^{\trans}$. The simulation reveals the coordinates of the attractive equilibrium point which correspond to the limit values of each coefficient.}
	\label{fig:aemodes}
\end{figure}

In the second example we consider the following uncertain dynamics studied in \citeasnoun{Iannelli2018}
\begin{align}\nonumber
\dot{x}_1 &= -x_2 - \frac{3}{2}x_1^2 - \frac{1}{2}x_1^3 + \spartt(\ranvar),\\\label{eq:vdpex2}
\dot{x}_2 &=  3x_1 - x_2 - x_2^2,
\end{align}
where $\spartt(\ranvar)\sim \text{Unif}(0.9,1.1)$  is a random variable depending on the stochastic germ $\ranvar \sim \text{Unif}(-1,1)$. This system has two equilibrium points whose location is uncertainty-dependent, and of which one is stable and the other unstable. Using the Legendre polynomial basis, we expand the system similarly to the first example and simulate a sample trajectory of the PC expanded system in order to determine the significant modes as well as the exact location of the stable equilibrium point. As Figure \ref{fig:aemodes} reveals choosing $\p = 2$ captures the significant modes. 
The stable equilibrium point lies at $\cxep =$ [0.4086, 0.7145, 0.0369, 0.0456, -4.9635e-04, -0.0012]$^{\trans}$. 

\begin{figure}[h]
	\centering
	\includegraphics[width=1\columnwidth]{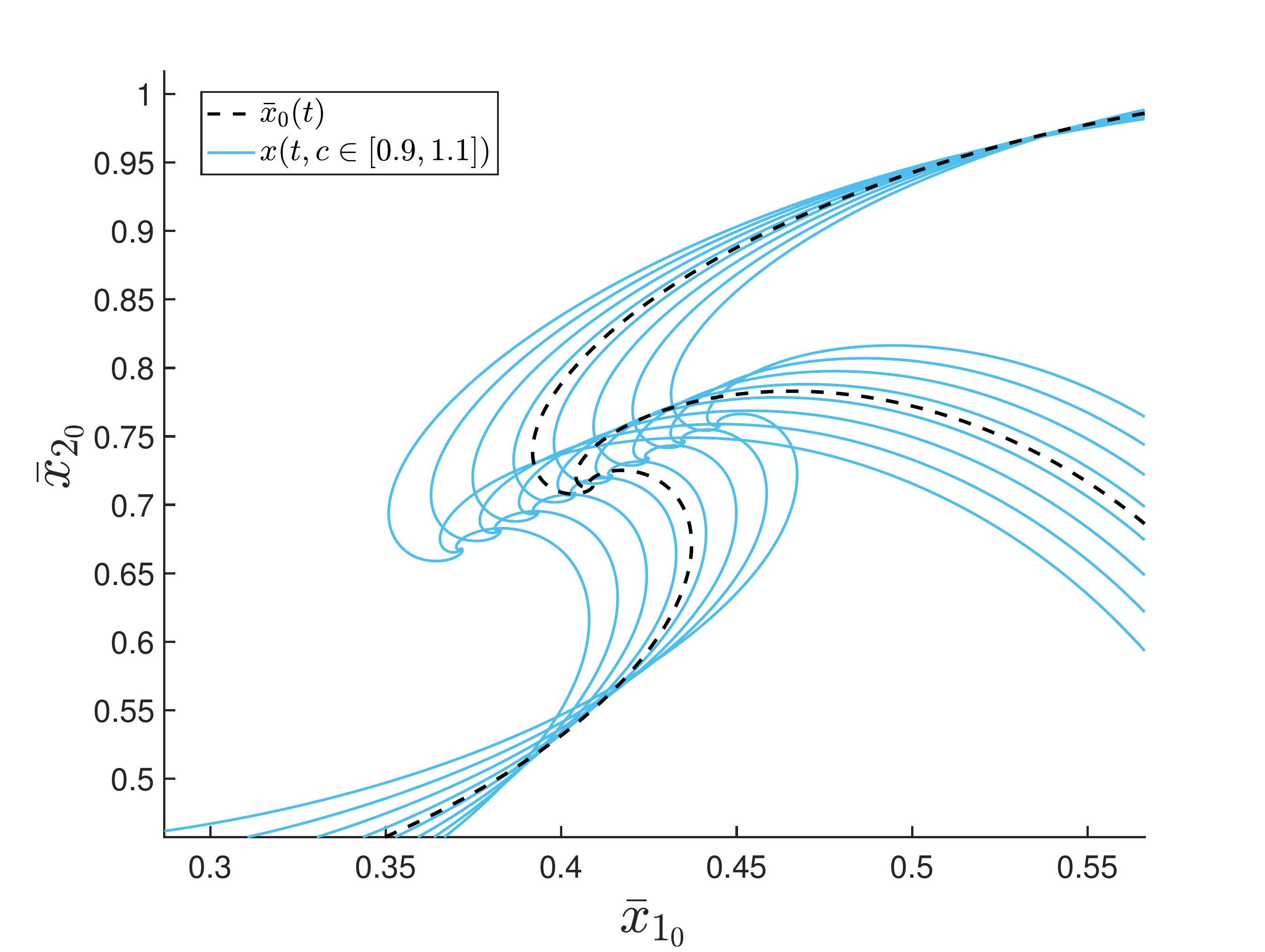}
	\caption{Depending on the realization of uncertainty the trajectories of the system \eqref{eq:vdpex2} converge to different equilibrium points, which together build the equilibrium set as given by Lemma \ref{lem:eqpoint}. The mean modes $\cx_0$ converge to the same point which corresponds to the equilibrium point of the PC expanded system.}
	\label{fig:aesamplesimuclose}
\end{figure}

As the PC expanded system has a non-zero equilibrium point, we obtain from Lemma \ref{lem:eqpoint} the equilibrium set $\invset$ of the stochastic system, which consists of all elements belonging to the distribution with mean $\mean = [0.408586, 0.7145229]^{\trans}$ and covariance $\varvar $= [4.533e-04, 5.603e-04; 5.603e-04, 6.923e-04]. Figure \ref{fig:aesamplesimuclose} illustrates this set by showing how trajectories from three different initial states converge to a different equilibrium point for different values of uncertainty. The yellow line indicates the $\cx_0$-trajectory of the PC expanded system which represents the mean of the stochastic system.

\begin{figure}[h]
	\centering
	\includegraphics[width=1\columnwidth]{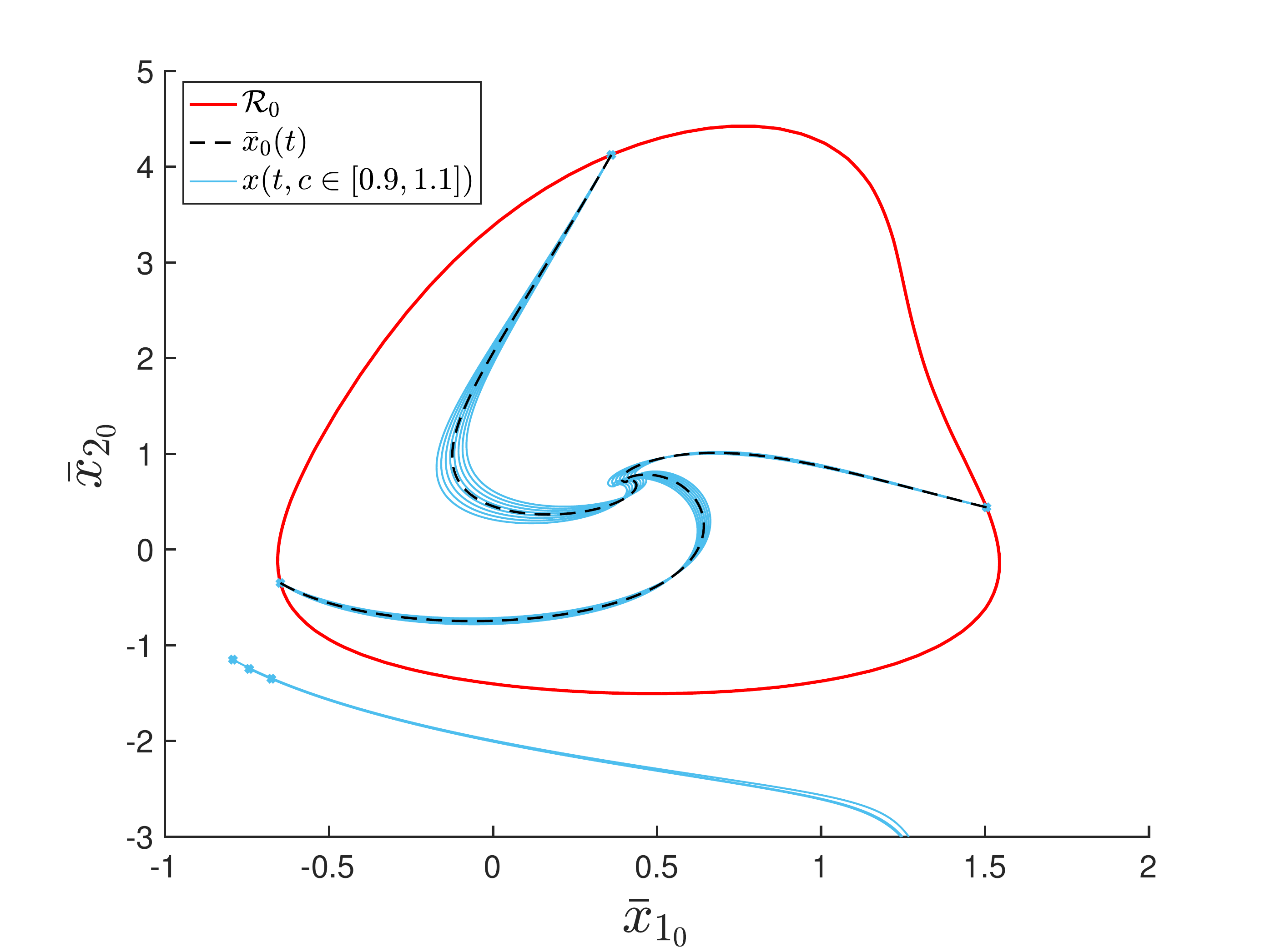}
	\caption{The result for $\roazero$ is shown in the red solid line. For various initial conditions on the boundary of $\roazero$ a Monte-Carlo simulation of the system \eqref{eq:vdpex2} for a range of realizations of the random variable is shown (blue lines). Note that the blue lines are each a trajectory of a deterministic system (obtained by sampling random values of the uncertainties in the stochastic dynamics); as for deterministic variables the mean equals the nominal value, $x_\text{nom} = \mean(x_\text{nom})$, the trajectories are plotted here in the mean coordinates $\cx_0$. The mean modes $\cx_0$ of the PC expanded system are also plotted (black dashed lines). For the three closest detected initial states with diverging trajectories the worst-case result of the Monte Carlo simulation is plotted. }
		%The diverging trajectory is plotted for illustration, showing how for some realizations of the uncertainty the system still converges, however for others and thus in the mean the trajectories diverge.}
	\label{fig:aesamplesimu}
\end{figure}

As in the first example, the ROA estimate is computed from \eqref{eq:finalprog} for a quartic $\lyp$ and the results used to obtain the ROA $\roazero$ in terms of the mean modes for zero variance on the initial state, as described in Remark \ref{rema:directrzero}. The results can be seen in Figure \ref{fig:aesamplesimu}. The comparison with the ROA estimates in \citeasnoun{Iannelli2018} shows that the approach proposed here provides similar sizes of ROA. To validate the results, we ran a Monte-Carlo simulation of the system \eqref{eq:vdpex2} for 1000 initial conditions on the boundary of the $\roazero$ for each of 20 realizations of the uncertainty ranging over the distribution. For illustration, three examples of the Monte-Carlo simulation using 8 realizations of the random variable over the distribution range are shown. The true ROA for this system is unknown. In order to obtain an upper bound on the conservatism of $\roazero$ we search for diverging trajectories by performing Monte Carlo simulations for a range of initial conditions located in the neighborhood outside of $\roazero$. The closest diverging trajectories found with the chosen sampling grid are shown in Figure \ref{fig:aesamplesimu}.

%The goal is to compute the smallest level set, obtained by uniformly expanding the certified inner estimate $\roa_0$, such that there is an initial condition on its boundary for which the system does not converge to the equilibrium. This is done by considering discrete values of the increasing level set size of $\roa_0$ (e$.$g$.$ 1.05, 1.1) and performing a Monte-Carlo search until escaping trajectories are detected. The closest diverging trajectories found with the chosen sampling grid are shown in Figure \ref{fig:aesamplesimu}.

%\begin{figure}[H]
%	\centering
%	\includegraphics[width=1\columnwidth]{figures/AE_roa_L2L4.eps}\vspace{-0.5cm}
%	\caption{Plot of $\roa_0$ for the case zero variance on the initial condition. The blue line shows the result obtained for $\deg(\lyp) = 2$ while the red line indicates the result for $\deg(\lyp) = 4$.}
%	\label{fig:aeroacases}
%\end{figure} 

\subsection{Comments on the numerical implementation}
The computational tractability of solving any SOS-program depends crucially on the size of the problem. The problem size scales exponentially in the number of states and polynomial degrees (polynomially, if scaled in either state or in polynomial degree alone). While the PC expansion approach does not alter the polynomial degrees it does lead to a $(\p+1)$-fold increase of the number of states. Depending on the number of modes needed to represent the system with a sufficient accuracy, the number of states can quickly become prohibitively large for low-dimensional stochastic systems. Research on more efficient SDP-solvers is ongoing and this limitation is likely to be alleviated in the future. One immediate remedy is offered by the DSOS/SDSOS framework introduced in \citeasnoun{Ahmadi2019}, which can solve SOS-programs tractably for up to 50 states. While potentially resulting in more conservative estimates these relaxations promise a significant speed up of the SOS program.

\section{Conclusion}
In this work we present a method to compute inner estimates of the region of attraction of stochastic nonlinear systems. The proposed method is applicable to a broad class of system consisting of second order processes which are affected by uncertainties coming from any $\ltwo$-distribution and which are further allowed to have uncertainty-dependent equilibria. The analysis is enabled by using Polynomial Chaos expansions through which a stochastic ODE is converted into a deterministic one. Using suitable stability notions in the form of moment boundedness and Lyapunov stability, it is shown how the ROA analysis of the PC expanded system offers direct information on the attractive behavior of the stochastic system for which a notion of a ROA is derived.
A numerical implementation for obtaining inner estimates of the ROA when the PC expanded system has a polynomial expression are provided via SOS optimization. The application to two examples taken from the literature shows that the proposed approach provides estimates of the ROA which are comparable to literature results obtained with less general methods. Further, the approach allows the user to obtain information on the stability of the system with defined statistical properties, such that if a particular uncertainty on the initial condition is known, the corresponding ROA estimate can be obtained.
The analysis method proposed here can be used and extended for various purposes among which are the stability analysis of systems with more complex equilibrium behavior, and the use of stochastic ROA analysis in controller design.

\bibliographystyle{autoarbib}

\bibliography{../../../../library} 





%\begin{thebibliography}{00}


%\bibitem{}

%\end{thebibliography}
\end{document}